\documentclass{emulateapj}
\usepackage{natbib,epsfig,lscape}
\def\lap{\lower.5ex\hbox{$\; \buildrel < \over \sim \;$}}
\def\gap{\lower.5ex\hbox{$\; \buildrel > \over \sim \;$}}

\def\ergcm2s{${\rm erg\ cm^{-2}\ s^{-1}}$}

\def\ergscm2s{${\rm erg\ cm^{-2}\  s^{-1}}$}

\def\cm-2{${\rm cm^{-2}}$}

\begin{document}

\title{The ACS Nearby Galaxy Survey Treasury I. The Star Formation History of the M81 Outer Disk}

\author{Benjamin F. Williams\altaffilmark{1}, 
Julianne J. Dalcanton\altaffilmark{1}, 
Anil C. Seth\altaffilmark{2},  
Daniel Weisz\altaffilmark{3}, 
Andrew Dolphin\altaffilmark{4}, 
Evan Skillman\altaffilmark{3}, 
Jason Harris\altaffilmark{5}, 
Jon Holtzman\altaffilmark{6}, 
L\'eo Girardi\altaffilmark{7}, 
Roelof S. de Jong\altaffilmark{8},
Knut Olsen\altaffilmark{9},
Andrew Cole\altaffilmark{10},
Carme Gallart\altaffilmark{11},
Stephanie M. Gogarten\altaffilmark{1}, 
Sebastian L. Hidalgo\altaffilmark{3},
Mario Mateo\altaffilmark{12}, 
Keith Rosema\altaffilmark{1}, 
Peter B. Stetson\altaffilmark{13}, 
Thomas Quinn\altaffilmark{1}
}

\altaffiltext{1}{Department of Astronomy, Box 351580, University of Washington, Seattle, WA 98195; ben@astro.washington.edu; jd@astro.washington.edu; stephanie@astro.washington.edu;  krosema@astro.washington.edu; trq@astro.washington.edu}
\altaffiltext{2}{CfA Fellow, Harvard-Smithsonian Center for Astrophysics, 60 Garden Street, Cambridge, MA 02138; aseth@cfa.harvard.edu}
\altaffiltext{3}{Department of Astronomy, University of Minnesota, 116 Church St. SE, Minneapolis, MN 55455; dweisz@astro.umn.edu; skillman@astro.umn.edu; slhidalgo@astro.umn.edu}
\altaffiltext{4}{Raytheon, 1151 E. Hermans Road, Tucson, AZ 85706; dolphin@raytheon.com}
\altaffiltext{5}{Steward Observatory, University of Arizona, 933 North Cherry Avenue, Tucson, AZ 85721; jharris@as.arizona.edu}
\altaffiltext{6}{Department of Astronomy, New Mexico State University, Box
30001, 1320 Frenger St., Las Cruces, NM 88003; holtz@nmsu.edu}
\altaffiltext{7}{Osservatorio Astronomico di Padova -- INAF, Vicolo
dell'Osservatorio 5, I-35122 Padova, Italy; leo.girardi@oapd.inaf.it}
\altaffiltext{8}{Space Telescope Science Institute, Baltimore, MD 21218; dejong@stsci.edu}
\altaffiltext{9}{NOAO, CTIO, Casilla 603, La Serena, Chile; kolsen@ctio.noao.edu}
\altaffiltext{10}{School of Mathematics and Physics, University of Tasmania, Hobart, Tasmania, Australia; andrew.cole@utas.edu.au}
\altaffiltext{11}{Instituto de Astrofísica de Canarias, Vía Láctea, s/n, 38200 La Laguna, Tenerife, SPAIN; carme@iac.es}
\altaffiltext{12}{Department of Astronomy, University of Michigan, 830 Denninson Building, Ann Arbor, MI 48109-1090; mmateo@umich.edu}
\altaffiltext{13}{Dominion Astrophysical Observatory, Herzberg Institute of Astrophysics,National Research Council, 5071 West Saanich Road, Victoria, BC  V9E 2E7, Canada; Peter.Stetson@nrc-cnrc.gc.ca}
  

\begin{abstract}

The ACS Nearby Galaxy Survey Treasury (ANGST) is a large {\it Hubble
Space Telescope (HST)} Advanced Camera for Surveys (ACS) treasury
program to obtain resolved stellar photometry for a volume-limited
sample of galaxies out to 4~Mpc.  As part of this program, we have
obtained deep ACS imaging of a field in the outer disk of the large
spiral galaxy M81.  The field contains the outskirts of a spiral arm
as well as an area containing no current star formation.  Our imaging
results in a color-magnitude diagram (CMD) reaching to m$_{F814W}$ =
28.8 and m$_{F606W}$ = 29.5, one magnitude fainter than the red clump.
Through detailed modeling of the full CMD, we quantify the age and
metallicity distribution of the stellar populations contained in the
field. The mean metallicity in the field is $-1<[{\rm M/H}]<0$ and
only a small fraction of stars have ages $\lap$1 Gyr. The results show
that most of the stars in this outer disk field were formed by $z\sim
1$ and that the arm structure at this radius has a lifetime of
$\gap$100 Myr.  We discuss the measured evolution of the M81 disk in
the context of surveys of high-redshift disk galaxies and deep stellar
photometry of other nearby galaxies.  All of these indicate that
massive spiral disks are mostly formed by z$\sim$1 and that they have
experienced rapid metal enrichment.

\end{abstract}

\section{Introduction}

Analytic and numerical models indicate that spiral disks should grow
and evolve with time due to continued gas accretion, interactions,
spiral density waves, and internal viscous evolution
\citep{fall1980,dalcanton1997,mo1998,vandenbosch2001,bell2002,shen2006,debattista2006,governato2007}.
Current observational constraints on the evolution of disks have come
largely through identifying changes in the bulk properties of the
galaxy population from in situ measurements at high redshifts (up to
$z\sim1-1.5$). However, these observational attempts to confirm disk
evolution models give conflicting results
\citep{simard1999,ravindranath2004,trujillo2004,barden2005,melbourne2007,cameron2007},
likely due to selection effects that are difficult to quantify.
Fortunately, the evolution of disks can be independently constrained
with photometry of the resolved stellar populations in nearby
galaxies.  Such photometry provides a fossil record of the formation
and evolution of the disk and complements the findings of redshift
surveys.

Resolved stellar photometry provides the most detailed data with which
to determine the star formation history (SFH) of a galaxy.  By fitting
stellar evolution models to an observed color-magnitude diagram (CMD),
we can recover the stellar ages and metallicities that best reproduce
the color and magnitude distribution of a galaxy's stars.  To fully
tap this capability, the ACS Nearby Galaxy Survey Treasury (ANGST) has
undertaken a program to measure resolved stellar photometry for a
volume-limited sample of galaxies \citep{dalcanton2008}. Within this
volume, large galaxies dominate the stellar mass. Of these, the
largest and most well-studied at all wavelengths is M81, a massive
SA(s)ab spiral disk at a distance of 3.9 Mpc \citep{tikhonov2005} with
low foreground extinction \citep[A$_V$=0.27; ][]{schlegel1998}.  Its
luminosity \citep[M$_K$=-24; ][]{skrutskie2006} places it at $2.5{\rm
L_*}$ \citep[assuming M$_{K*}$=-23; ][]{kochanek2001}, making it characteristic
of the disk galaxies seen in redshift surveys out to $z\sim 1$
\citep[e.g.,][]{oyaizu2008}.

M81 provides a key laboratory for using resolved stellar photometry to
look for relics of the evolution seen at high-redshift.
Several surveys suggest that large disk galaxies like M81 have their
disks in place by $z\sim 1$, after which luminosity evolution
dominates
\citep{lilly1998,ravindranath2004,papovich2005,sargent2007,melbourne2007}.
In this context, understanding the SFH of nearby large disk galaxies
like M81 provides complementary insight into their evolution that
is free of any biases contained in redshift survey samples.

Much work has been done to understand the evolution of M81 from its
star clusters \citep{ma2005,perelmuter1995}, its X-ray source
population \citep{swartz2003}, and its young supergiant stars as seen
in the near infrared \citep{davidge2006}.  Its structure and evolution
have been studied in detail from ultraviolet to radio
\citep{perez-gonzalez2006,willner2004,gordon2004,li2004,westpfahl1998},
but very little work has been done to integrate these results with the
resolved stellar populations.

Most work on resolved stellar populations in M81 has relied on the
{\it Hubble Space Telescope (HST)} to resolve the individual stars, as
only stars brighter than the red giant branch (RGB) tip are resolved
from the ground \citep[e.g.,][]{madore1993}.  \citet{tikhonov2005}
studied the resolved stellar populations of M81 with archival WFPC2
and ACS data, and there are several {\it HST} programs that are
currently underway with greater depth and spatial coverage
\citep{dejong2007,Sarajedini2005,zezas2005,huchra2004}.  The results
from these programs will allow detailed comparisons with those from
integrated light studies.

Herein we describe the ANGST measurements of the SFH of the outer disk
of M81 using F606W and F814W stellar photometry.  The outer disk has
low enough crowding that precision photometry can be obtained below
the red clump with ACS, and provides leverage with which to
potentially constrain size evolution of the galactic disk.  \S2
details our data acquisition and reduction techniques. \S3 discusses
our analysis of specific portions of the color-magnitude diagram
(CMD), and \S4 details our analysis of the full CMD to determine the
SFH of the field.  Finally, \S5 summarizes the conclusions drawn from
these results, placing our measurements in the context of disk galaxy
surveys and resolved stellar populations of other large galaxies.  We
adopt a cosmology with $H=73$ km s$^{-1}$ Mpc$^{-1}$, $\Omega_\Lambda
= 0.76$, and $\Omega_M = 0.24$ for all lookback time calculations, and
we assume a distance to M81 of 3.9 Mpc for conversions of angular
measurements to physical distances.

\section{Data Acquisition and Reduction}

\begin{figure}
\centerline{\epsfig{file=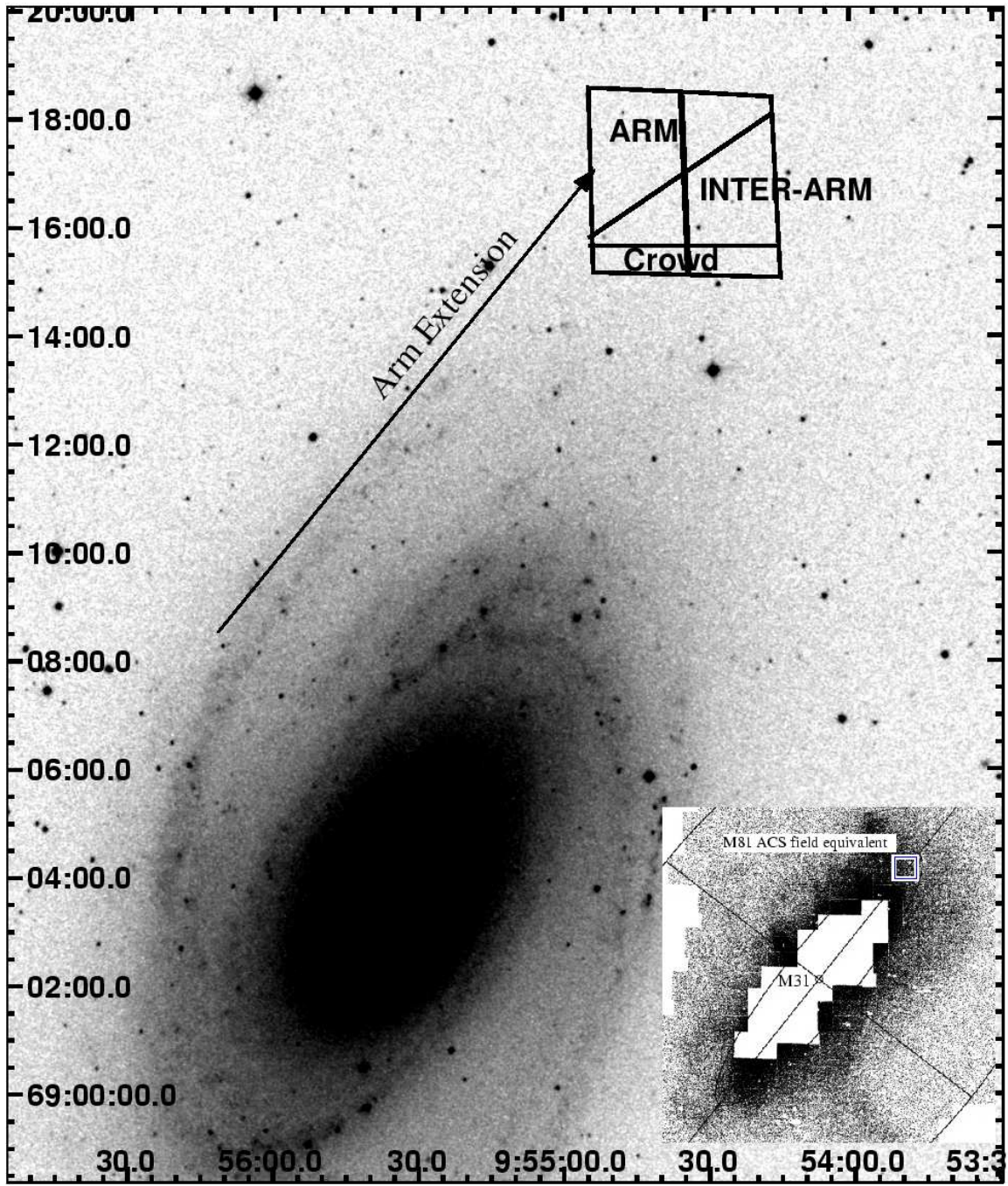,width=6.0in,angle=0}}
\caption{The locations of our M81-DEEP field and our defined
subregions of the field shown on a DSS image.  The arrow marks the
apparent spur of the Northern spiral arm according to the distribution
of main-sequence stars in our field. Inset on the lower-right corner
is the equivalent location of our M81-DEEP field shown with a white
box on a star count map of M31 \citep{ferguson2002}.  The inner and
outer edges of the field are located at the same scale lengths as in
M81.}
\label{field_loc}
\end{figure}

From 2006-Nov-16 to 2006-Nov-22, we observed a field in the outskirts
of the M81 disk located at R.A.~(2000) =148.644625 (09:54:34.7),
decl.~(2000) = +69.2804 (+69:16:49.4) with a rotation angle
PA\_V3=89.81.  Figure~\ref{field_loc} shows an outline of the field
location, which is 14$'$ (16 kpc at M81) out along the major axis and
corresponds to 5 scale lengths an 18 effective radii of the bulge
\citep[$h_r \sim 3$ kpc and bulge $R_e \sim 0.9$ kpc,][]{kendall2008}.
The equivalent location in M31, as shown on the inset in
Figure~\ref{field_loc}, suggests that the disk population is likely to
still dominate at this radius if M81 and M31 are similar.  This
suggestion is consistent with the ongoing analysis of
\citet{dejong2007}, which finds that the halo population does not
dominate until galactocentric distances $\gap20$ kpc.

%NED: morph type: SA(s)ab
%NED; min/maj = 0.524; inc = 58.4 deg

We obtained 9 full-orbit exposures with the ACS \citep{ford1998}
through the F606W (wide $V$) filter, and 11 full-orbit exposures
through the F814W ($I$ equivalent) filter.  These data totaled 24132~s
and 29853~s of exposure time in F606W and F814W, respectively.
Routine image calibration, including bias corrections and
flat-fielding, were performed by the {\it HST} pipeline, OPUS version
2006\_5, CALACS version 4.6.1.  We processed the images by running the
{\tt multidrizzle} task within PyRAF \citep{koekemoer}. This procedure
was used only to flag the cosmic rays in the individual images, after
which, photometry was measured simultaneously for all of the objects
in the uncombined images using the software package DOLPHOT
\citep{dolphin2000} including the ACS module.  This package is
optimized for measuring photometry of stars on ACS images using the
well-characterized and stable point spread function (PSF), calculated
with TinyTim.\footnote{http://www.stsci.edu/software/tinytim/} The
software fits the PSF to all of the stars in each individual frame to
find PSF magnitudes.  It then determines and applies the aperture
correction for each image using the most isolated stars, corrects for
the charge transfer efficiency of the ACS detector, combines the
results from the individual exposures, and converts the measured count
rates to the VEGAmag system.

\begin{figure}
\centerline{\epsfig{file=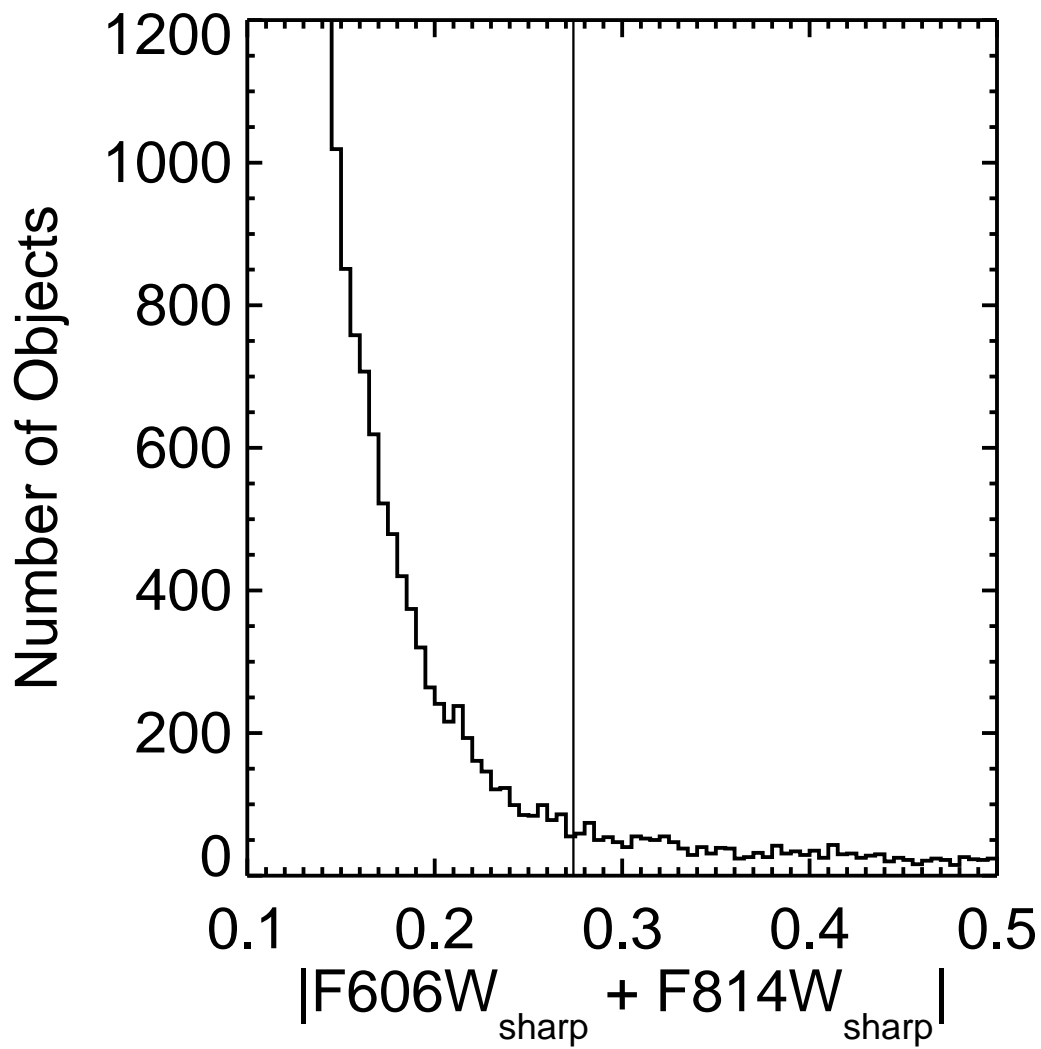,width=6.0in,angle=0}}
\caption{{\it Histogram}: The distribution of the combined sharpness
values for objects in our initial photometry catalog. {\it Vertical
Line}: The sharpness cut we applied to our final photometry catalog.}
\label{sharp}
\end{figure}

\begin{figure}
\centerline{\epsfig{file=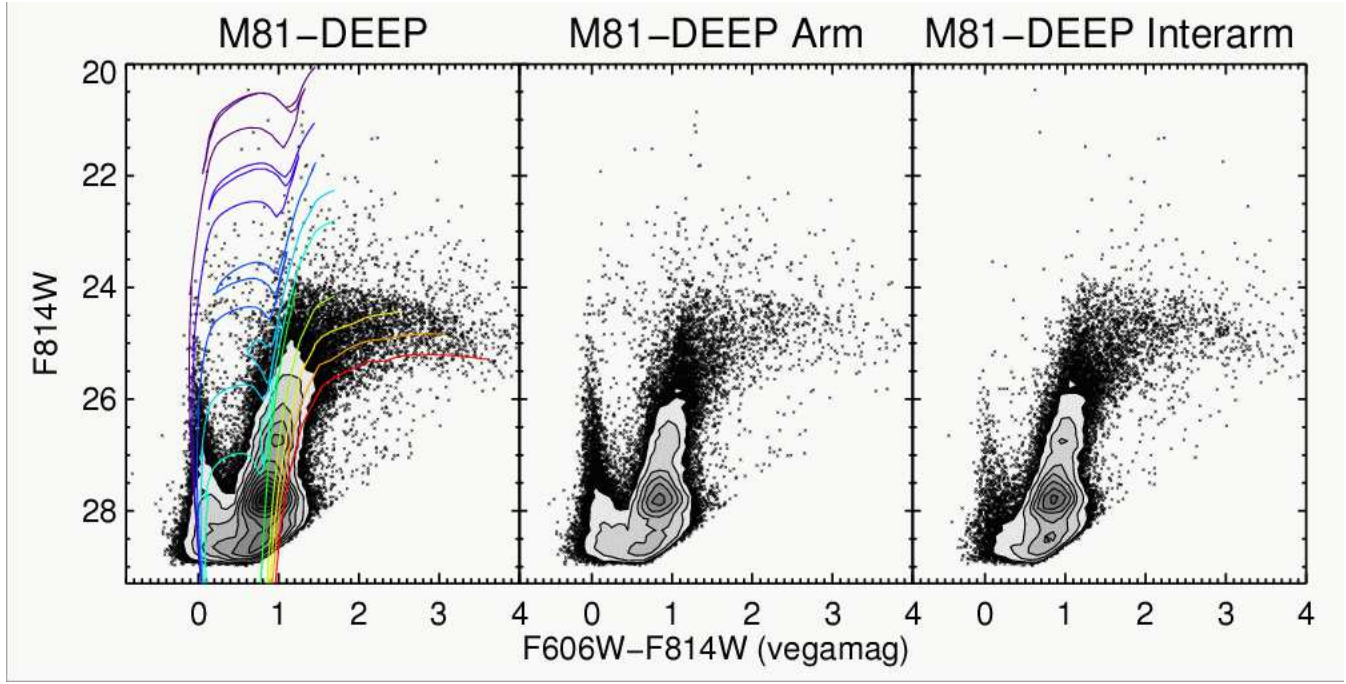,height=7.0in,angle=270}}
\caption{{\it Left:} The F606W, F814W CMD for our entire ACS
field. Lines show a small subset of the isochrones used to fit the
data \citep[][shifted assuming $A_V=0.3$ and $(m-M)_0 =
27.9$]{marigo2008}.  Isochrones shown are (from blue to red):
[M/H]=-0.4, log(age)=7.3; [M/H]=-0.4, log(age)=7.6; [M/H]=-0.4,
log(age)=8.0; [M/H]=-0.4, log(age)=8.3; [M/H]=-0.4, log(age)=8.6;
[M/H]=-1.3, log(age)=10; [M/H]=-0.7, log(age)=10; [M/H]=-0.4,
log(age)=10; [M/H]=-0.2, log(age)=10; [M/H]=0, log(age)=10. {\it
Middle:} The CMD of the arm region shown in Figure~\ref{subdivide}.
{\it Right:} The CMD of the interarm region shown in
Figure~\ref{subdivide}. In areas where the points would saturate the
plot, we provide contours following the density of points in that part
of the CMD.}
\label{cmds}
\end{figure}

The DOLPHOT output was then filtered to only allow objects classified
as stars with signal-to-noise $>$6 in both filters.  The list was
further culled using sharpness ($|F606W_{sharp} + F814W_{sharp}| <
0.27$) and crowding ($F606W_{crowd} + F814W_{crowd} < 0.1$).  The
sharpness cut was chosen based on the distribution of values in the
original catalog.  This distribution is shown in Figure~\ref{sharp}
and flattens at a value of $\sim$0.27.  The crowding parameter gives
the difference between the magnitude of a star measured before and
after subtracting the neighboring stars in the image.  When this value
is large, it suggests that the star's photometry was significantly
affected by crowding, and we therefore exclude it from our catalog.
Quality cuts based on the $\chi$ values were also considered, but they
were rejected when a correlation was found between $\chi$ and the
local background.  Our final star catalog for the field contained
120912 stars detected in both F606W and F814W, and the resulting CMD
is shown in Figure~\ref{cmds}.

The same software package was used to perform artificial star tests
using identical measurement techniques and quality cuts.  We ran
2.5$\times$10$^6$ artificial stars to characterize our photometry
errors and completeness as a function of color, magnitude, and
position.  In each iteration, a single star was added to the images,
and the photometry of the images was remeasured in the area where the
star was added, including a radius of the PSF size plus the photometry
aperture size to include the photometry of all stars whose photometry
could be affected by the existence of the artificial star.  Half of
the artificial stars were sampled randomly in color and magnitude
covering the full range present in our observed CMD plus an additional
magnitude fainter to account for upscatter of faint stars into our
recovered magnitude range.  The other half were sampled following the
color and magnitude distribution of our observed CMD after
extrapolating the distribution to fainter magnitudes to account for
upscatter.  The artificial stars were distributed randomly over the
field of view.  The photometric errors measured from our tests are
shown in Figure~\ref{ast_errs}, and the completeness measured from our
tests is shown in Figure~\ref{ast_comp}. When fitting the
color-magnitude distribution of the stars, we included only stars
brighter than a 60\% completeness limit as measured from the
artificial star tests ($m_{F606W} = 29.1$ and $m_{F814W} = 28.4$).  At
this depth and this Galactic latitude ($b = 40.9^{\circ}$), the
expected number of Galactic foreground stars is $\lap$20
arcmin$^{-2}$, suggesting foreground contamination in our field of
$\lap$200 objects or $<$0.2\% of the total number of stars.

\section{Red Clump and Asymptotic Giant Branch Bump Analysis}\label{rcfit}

\subsection{Overview}

Before attempting to understand the complexities of the CMD as a
whole, we first focus on a few key stellar evolution features in our
full-field CMD that help to give a broad sense of the range of ages
and metallicities of the stars in the field (Figure~\ref{cmds}).  The
vertical plume at $m_{F606W}-m_{F814W}=0$ is due to the young
population of stars that is still on the main sequence.  The handful
of stars populating the diagram brightward and slightly redward of
this plume are massive core He-burning stars, which are young stars
that have recently evolved off of the main sequence.  We plot the
spatial distribution of the main sequence stars in
Figure~\ref{subdivide}; the young stars are clearly concentrated in
the inner disk (seen in the bottom of the image) and in the extension
of the spiral arm seen in Figure~\ref{field_loc}.

Several older populations are seen as well.  The dense clump of stars
at $m_{F814W}\sim27.8$ is the red clump
\citep{cannon1970,sarajedini1999}, which corresponds to the core-He
burning phase of all intermediate-age and old populations which are
neither too old, nor too metal-poor, to develop the horizontal branch.
Brightward of this feature is another, less prominent peak, known as
the asymptotic giant branch (AGB) bump \citep{gallart1998}, which
corresponds to the so-called early-AGB phase of low-mass stars, during
which He shell burning transits from a very extended to a thin shell.
Extending vertically through both the red clump and AGB bump is a
broad RGB with a well-defined tip, indicating the presence of a range
of ages and metallicities in the field.  Brightward of this tip is a
relatively small number of thermally-pulsating AGB stars, i.e. stars
in the stage of double shell burning which undergo recurrent He shell
flashes. The most luminous among these thermally-pulsating AGB stars
have intense winds that eventually shed their outer layers to become
planetary nebulae.

\subsection{Red Clump and AGB Bump Fitting Method}

To investigate the general characteristics of the stellar populations
in our field, we first performed precision measurements of the very
well-defined red clump and AGB bump, seen at $m_{F814W}\sim27.8$ and
$m_{F814W}\sim26.7$ in the diagrams in Figure~\ref{cmds} respectively.
These are areas of the CMD containing a high density of stars,
allowing their precise locations to be measured and directly compared
to stellar evolution models.  The colors and luminosities of these
features have been used to constrain the properties of stellar
populations \citep{rejkuba2005,tanaka2008}.  We include such a
measurement here to take advantage of the depth and quality of our
photometry and to allow for intercomparisons with other such
analyses. However, as we discuss below, uncertainties with the models
of these discrete features substantially limit the interpretation of
these measurements, making simultaneous fitting of the full CMD
preferable; we present such fits in \S~4.

We measured the magnitudes of the red clump and AGB bump by creating a
magnitude histogram from our photometry and fitting it with a
combination of a line and 2 Gaussians, following the methods of
\citet{rejkuba2005}. We then fit Gaussians to histograms of slices
through color space taken at the measured peak magnitudes
(Figure~\ref{fits}).  The red clump in this outer region of the M81
disk has $m_{F814W}=27.792\pm0.002$ (FWHM=0.450) and $m_{F606W} -
m_{F814W}=0.855\pm0.001$ (FWHM=0.363).  The AGB bump has
$m_{F814W}=26.72\pm0.01$ (FWHM=0.27) and $m_{F606W} -
m_{F814W}=0.974\pm0.002$ (FWHM=0.346).  These are the raw measurements
for these features; no extinction correction has been applied.

We then measured the locations of model red clumps and AGB bumps by
populating the isochrones of \citet{girardi2002} including updates of
AGB models in
\citet{marigo2008}.\footnote{http://stev.oapd.inaf.it/cmd}
We used the StarFISH \citep{harris2001} task {\tt testpop}, to produce
a grid of CMDs for discrete ages and metallicities, assigning
extinctions, distances, errors, and completeness measured from our
data (see \S~\ref{sfh}).  We then ran our red clump and AGB bump
fitting technique on this grid of model CMDs to measure the location
of these features as a function of age and metallicity.

We determined the model red clump and AGB bump with F606W and F814W
magnitudes closest to those observed in our field, assuming an adopted
distance of $(m-M)_0=27.93$ and extinction of $A_V=0.27$.  These
distance and extinction values resulted in a grid of model red clump
and AGB bump values that did not intersect with our measured values.
We therefore show a model grid with a farther assumed distance
($(m-M)_0=28.0$) and higher extinction ($A_V=0.3$) in
Figure~\ref{rejkuba}, to produce a grid that does intersect with our
observed values.  From this analysis, the models suggest that the red
clump is dominated by stars with an age of $\sim$2--3 Gyr and
[M/H]$\sim$-0.4 and that the AGB bump is dominated by older, more
metal-poor stars with an age of $\sim$10 Gyr and [M/H]$\sim$-0.7.

Because the precise colors and magnitudes of the red clump and AGB
bump do not provide reliable conclusions (see S~\ref{interpretation}),
we also investigated the relative numbers of stars contained within
the red clump and AGB bump features.  This measurement should be
insensitive to distance and extinction.  We calculated the integral of
the functional fit to the magnitude histogram including and excluding
the Gaussian terms corresponding to each feature.  The difference
between these integrals provided the number of stars within each
feature.  The ratio of the number of stars in the AGB bump to that
contained in the red clump ($N_{AGBb}:N_{RC}$) was 0.044$\pm$0.008.
Errors are the standard deviation of the same measurement made on 100
random samples of our data and are dominated by the small number of
stars in the AGB bump.  Comparing this range of ratios to the ratios
obtained by running the same calculations on the functional fits to
models suggests that the dominant stellar population of our sample is
metal-rich and older than $3$~Gyr (see Figure~\ref{ratios}).  The
ratio is therefore consistent with the color and magnitude of the
model AGB, but less so with those of the red clump, in that the model
red clump is fainter than it should be for a population old enough to
have this fraction of AGB bump stars.

\subsection{Interpretation of Red Clump and AGB Bump Fitting Results}\label{interpretation}

Although analyses similar to those above are becoming common, the red
clump and AGB-bump colors, magnitudes, and relative numbers cannot
provide conclusive results about the age and metallicity distribution
of the population.  The reasons for the inconclusive results include
the sensitivity of the results to known deficiencies of stellar
evolution models as well as to the assumed distance and reddening of
the stars in the CMD.

We found our fits to the color and magnitude of the red clump and AGB
bump difficult to interpret because they are particularly sensitive to
weaknesses of the stellar evolution models.  The majority of the red
clump models are faint compared to our observed red clump.  While the
trends of the colors and luminosities of the red clump and AGB bump
with age and metallicity (older and more metal-rich are fainter and
redder) are robust against model updates and different model
prescriptions, the absolute colors and luminosities are less stable.
For example, red clump model luminosities depend primarily on
opacities and neutrino losses during the previous RGB phase, which are
still subject to significant variations in the
literature. \citet{girardi2002} models of the red clump are known to
be some of the faintest because of their treatment of these processes
\citep{castellani2000}. This effect can be seen in
Figure~\ref{rejkuba}, where nearly all of the model red clumps are
fainter than our observed red clump even though most of the model AGB
bumps are brighter than our observed AGB bump.

The $N_{AGBb}:N_{RC}$ ratio is also difficult to interpret.  The ratio
predicted by the models is sensitive to the assumed helium content and
the treatment of convective cores, including complex processes such as
overshooting and semiconvection, rendering any conclusions based on
this ratio alone unreliable.

Furthermore, fits to the color and magnitude of the red clump and AGB
bump depend on the extinction and distance values applied to the
models, as changes of only $\sim$0.1 mag in color or luminosity of the
features correspond to differences of $\sim$0.3 dex in metallicity and
log(age).  The effects of distance and reddening uncertainty are shown
with the arrows in Figure~\ref{rejkuba}.  While the adopted distance
affects mainly the measured age, the adopted reddening largely affects
the measured metallicity.  More specifically, if we assume only
foreground extinction value ($A_V=0.27$), our best-fitting ages remain
the same, but the metallicities become [M/H]$\sim$-0.4 and
[M/H]$\sim$-0.1 for the AGB bump and red clump, respectively. If we
assume a distance modulus of $(m-M)_0=27.9$, the best-fitting ages and
metallicities are [M/H]$\sim$-0.7, [M/H]$\sim$-0.5 and $\sim$13 Gyr,
$\sim$5 Gyr for the AGB bump and red clump respectively.

In summary, while studying the red clump and AGB bump properties in
detail is helpful for getting some sense of the overall age and
metallicity of the population, or perhaps for constraining relative
ages and metallicities among different galaxies, the results'
sensitivity to uncertainties in stellar models, distance, and
extinction, make this analysis less than optimal for obtaining
reliable information about the stellar populations.  Furthermore, no
single age and metallicity will precisely fit the observed values
because these features contain populations of a range of ages and
metallicities.  These difficulties with single component fitting make
it necessary to perform more sophisticated statistical fits to the
entire CMD to decipher the range of ages and metallicities present in
our field. Full CMD fitting helps to reduce the effects of model
deficiencies, providing an overall picture of the age and metallicity
of the population even if some details of the models are wrong.

\section{Full CMD Fitting}\label{sfh}

\subsection{Fitting Technique and Models Used}

We measured the complete SFH, including the star formation rate and
metallicity as functions of age, using the MATCH package
\citep{dolphin2002}. This software fits the entire observed CMD by
populating the stellar evolution models of \citet[][with updated AGB
models from \citealp{marigo2008}]{girardi2002} with a given initial
mass function (IMF), finding the distance modulus, extinction, and
linear combination of ages and metallicities that best fit the
observed color and magnitude distribution \citep[see details
in][]{dolphin2002}.

As was the case for the simplified AGB bump and red clump analysis,
the results of full CMD fitting rely on the stellar evolution models
used to fit the CMD.  There are known discrepancies between different
sets of stellar evolution models in the RGB, red clump, and AGB bump
phases of evolution (see Figure 10 of \citealp{gallart2005a}). The
reasons for the discrepancies are many, from the adopted input physics
to the uncertainty in mass loss during the RGB and AGB phases of
evolution.  These differences affect the location and morphology of
these regions of the observed CMD.

On the other hand, the shape of the red clump region does contain
information about the SFH.  The fitting algorithms are sensitive to
this information if they are set to fit the distance and reddening
values independently in order to compensate for offsets between model
red clump positions. For instance, the RGB colors are sensitive to
both age and metallicity, whereas the RGB-red clump color difference
is more sensitive to the age \citep{hatzidimitriou1991}. While many
age-metallicity models could fit the overall RGB color distribution,
when the red clump is present its color difference forces the
SFH-recovery towards the best-fitting age distribution, provided that
it is allowed to fit also the red clump magnitude (e.g. by adjusting
the distance). The final result is that this area of the CMD provides
a reliable SFH, even if the models for the red clump contain offsets.
To test this point, we ran a different fitting code and stellar
evolution library to fit our full-field data. We applied the
IAC-STAR/IAC-POP/MINNIAC CMD fitting codes \citep[][Hidalgo, S. L. et
al. 2008, AJ, in preparation]{aparicio2004,aparicio2008} to our data
using the BaSTI stellar evolution models \citep{pietrinferni2004}.
The broad trends on the resulting SFH were totally consistent.

This point is strengthened by the tests of \citet{barker2007}, who
compared results for SFHs as determined with different sets of stellar
evolution models and found differences in the details but agreement
for general population characteristics.  Furthermore, the results of
\citet{tolstoy1998} and \citet{cole2007} show that, while SFHs based
on the main-sequence turnoff provide more reliable age information at
older ages than those from shallower photometry, the general trends
determined in shallower photometry are robust.  To avoid drawing
conclusions based on the finest details in age and metallicity applied
to fit the CMD (see \S~\ref{binning}), we bin our results to an age
and metallicity resolution where both the CMD fitting method and
stellar evolution models are well-tested and the limitations are
known.

\subsection{Field Division}

\begin{figure}
\centerline{\epsfig{file=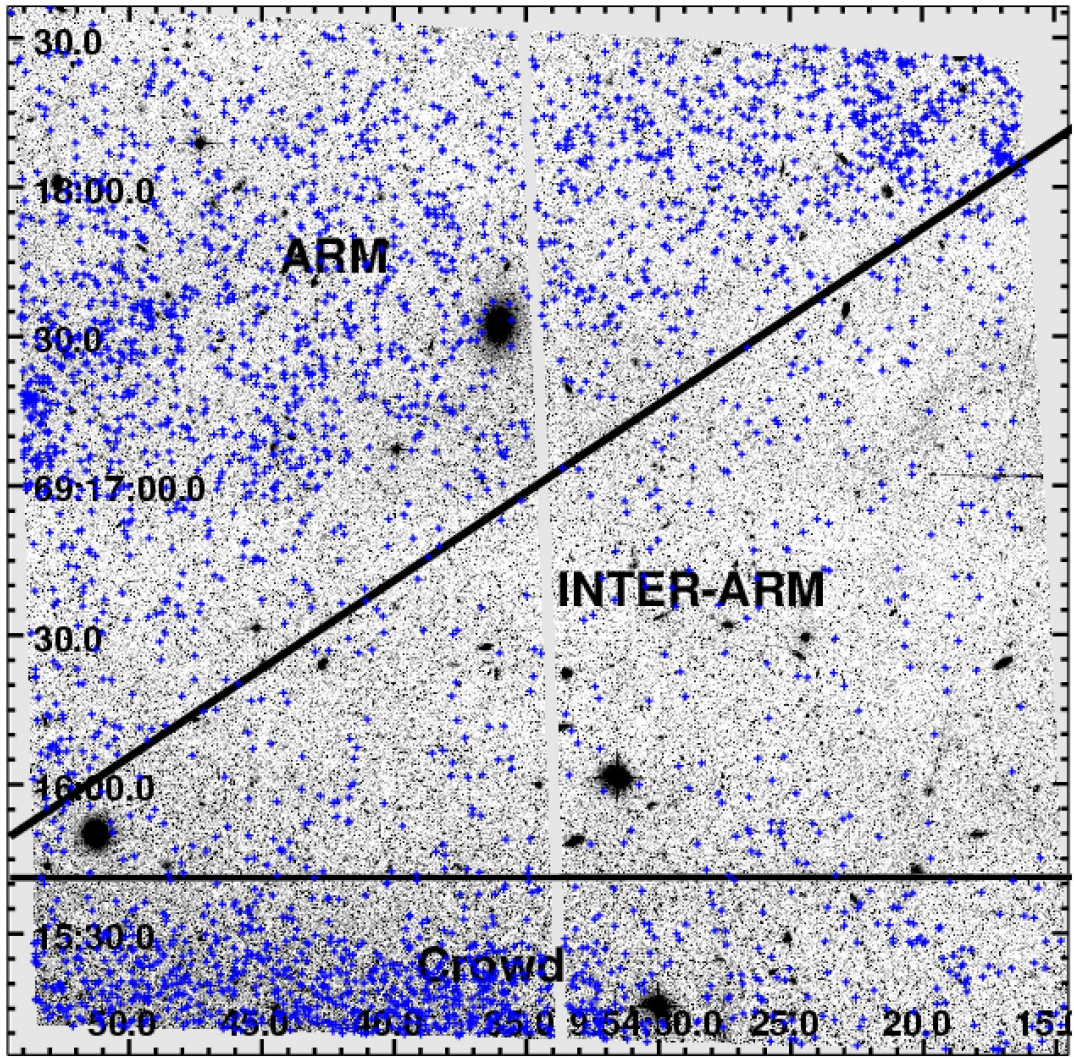,width=6.0in,angle=0}}
\caption{The borders of the subregions used in our SFH analysis.  Blue
crosses mark the locations of main sequence stars in the field.  Main
sequence stars were chosen using a hand-drawn polygon that followed
the edges of the blue plume of stars in the CMD.  The color and
magnitude limits of the polygon were approximately 24$<$F814W$<$28 and
-0.2$<$F606W-F814W$<$0.3.}
\label{subdivide}
\end{figure}

\begin{figure}
\centerline{\epsfig{file=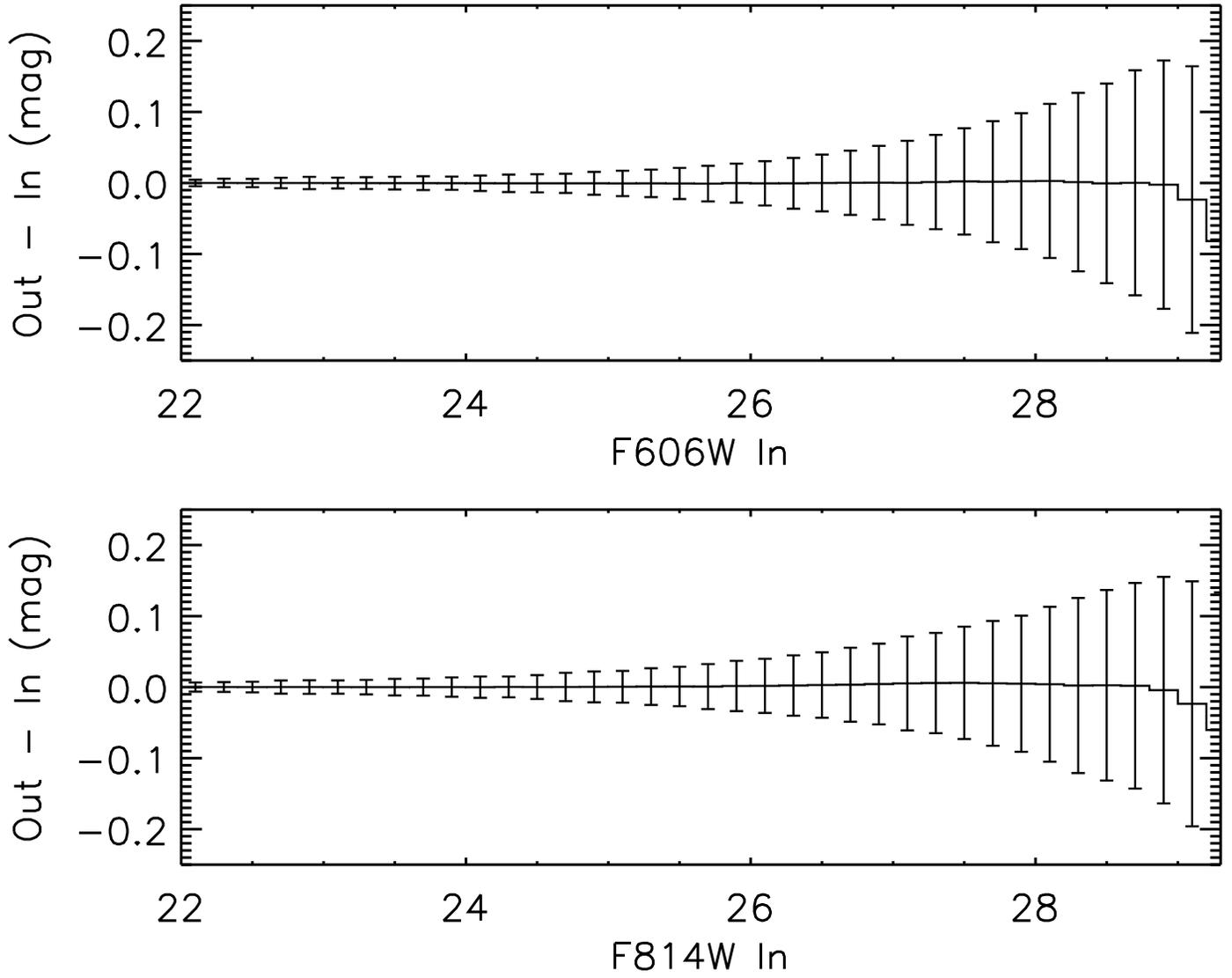,width=6.0in,angle=90}}
\caption{{\it Top:} The mean residual magnitude and root-mean-square
error of the artificial star tests in the F606W filter measurements
are shown as a function of input star magnitude.  {\it Bottom:} Same
as {\it Top}, but for the F814W filter measurements.}
\label{ast_errs}
\end{figure}

\begin{figure}
\centerline{\epsfig{file=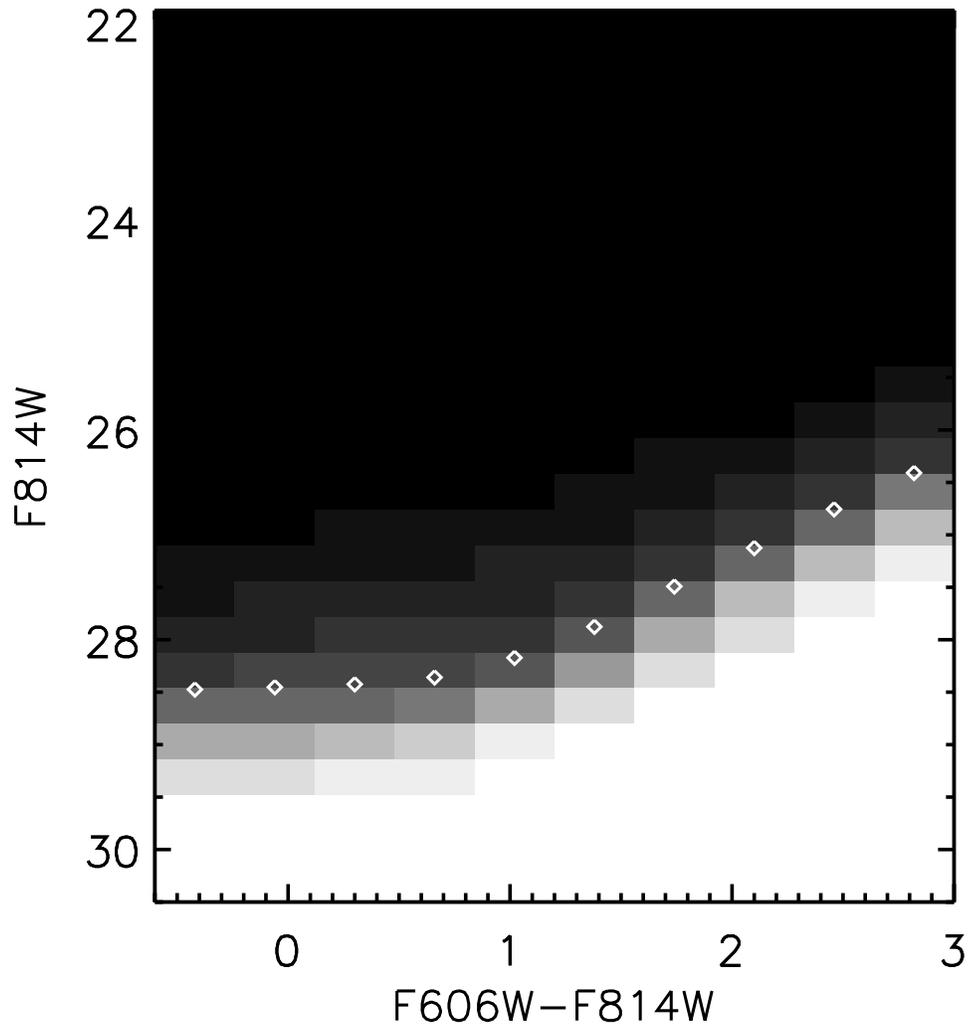,width=6.0in,angle=0}}
\caption{The completeness measured from our artificial star tests is
shown in grayscale as a function of color and magnitude.  The scale is
linear, with 100\% completeness shown as black and 0\% completeness
shown as white. White diamonds mark the completeness limit used to fit
the SFH of the field.}
\label{ast_comp}
\end{figure}

\begin{figure}
\centerline{\epsfig{file=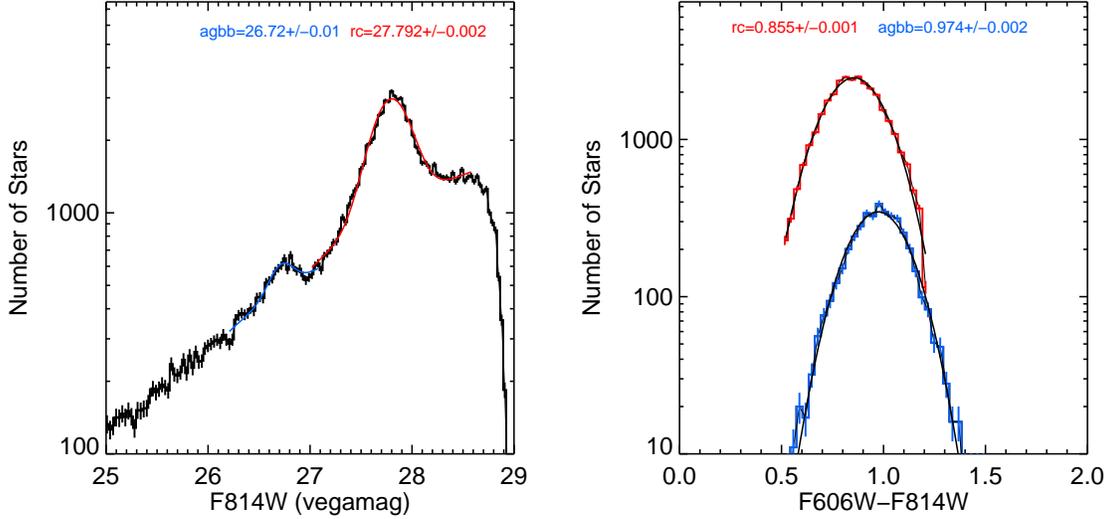,width=6.0in,angle=0}}
\caption{{\it Left:} Gaussian plus line fits to our measured magnitude
histogram to determine the magnitude of the red clump (drawn in red)
and AGB bump (drawn in blue).  {\it Right:} Gaussian fits to the
color histograms measured at the best-fit magnitude of the red clump
(red) and AGB bump (blue).}
\label{fits}
\end{figure}

\begin{figure}
\centerline{\epsfig{file=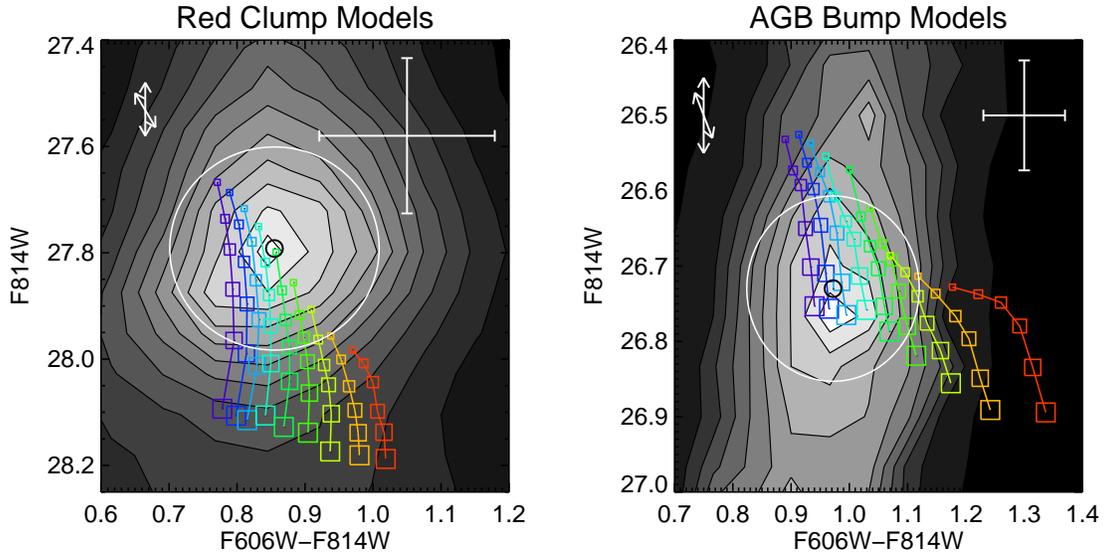,width=6.0in,angle=0}}
\caption{{\it Left:} Small portion of our full-field CMD centered on
the red clump region.  The black circle marks the best-fitting center
for the feature. The white circle marks the measured height and width
of the feature ($\sigma_{({\rm m}_{F606W}-{\rm m}_{F814W})}$ = 0.154;
$\sigma_{{\rm m}_{F814W}}$ = 0.191).  Boxes mark the locations of the
fits to the same feature in model CMDs convolved with our photometric
errors and completeness.  Redder colors denote higher metallicities
(the metallicity range is -1.0$<$[M/H]$<$0.2); larger boxes denote
older ages (2 Gyr$<$age$<$13 Gyr).  Large error bars in the upper
right show the 1-sigma width of the feature in a single age and
metallicity CMD ($\sigma_{({\rm m}_{F606W}-{\rm m}_{F814W})}$ = 0.129;
$\sigma_{{\rm m}_{F814W}}$ = 0.146).  Arrows show our 1-sigma error
ranges for distance and reddening.  ({\it Right:} Same as {\it Left},
but for a small portion of our full-field CMD centered on the AGB bump
region.  The measured height and width of the feature are
$\sigma_{({\rm m}_{F606W}-{\rm m}_{F814W})}$ = 0.148 and $\sigma_{{\rm
m}_{F814W}}$ = 0.123.  Values for the 1-sigma width of the feature in
the a single age and metallicity CMD are $\sigma_{({\rm
m}_{F606W}-{\rm m}_{F814W})}$ = 0.070 and $\sigma_{{\rm m}_{F814W}}$ =
0.073) Note the fitted AGB bump center is slightly brighter than the
mode value due the the slope of the luminosity function skewing the
mode to fainter magnitudes. Predicted magnitudes assume $A_V=0.3$ and
$(m-M)_0=28.0$.}
\label{rejkuba}
\end{figure}

\begin{figure}
\centerline{\epsfig{file=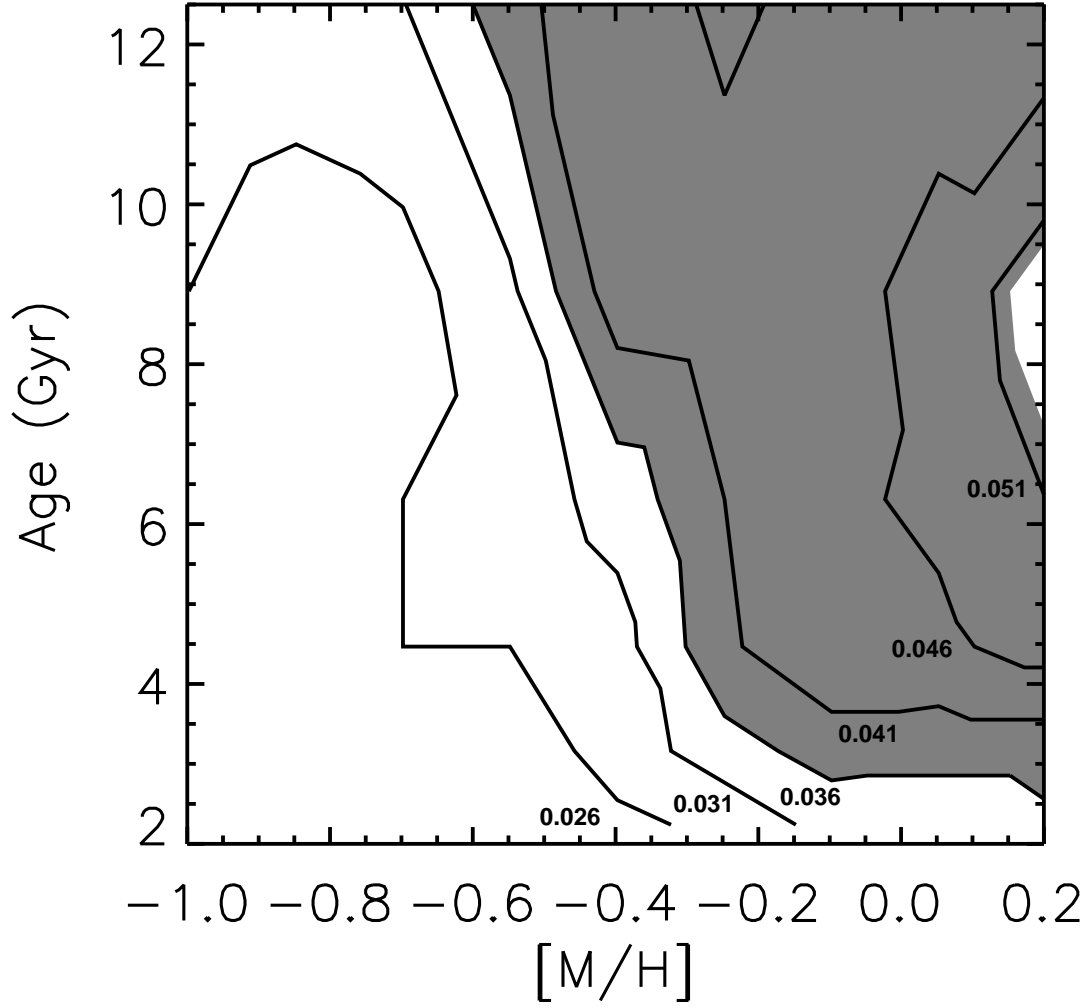,width=6.0in,angle=0}}
\caption{The age-metallicity plane.  Contours denote different
$N_{AGBb}:N_{RC}$ ratios, increasing from 0.026 to 0.051 from left to
right and labeled.  The shaded area denotes where model stellar populations have a
ratio $N_{AGBb}:N_{RC}$ consistent with that measured from our data.}
\label{ratios}
\end{figure}

Our field contained a portion of a spiral arm running through the
north and east portion of the image.  The bifurcated arm, which may
have been created by an interaction, can be seen in the locations of
the main-sequence stars, which are mainly limited to the northeast
portion and the southeast corner (Figure~\ref{subdivide}).  This
distribution shows that the northeast spiral arm is split, with a spur
to the north, consistent with H~I maps of \citet{yun1994} and
\citet{adler1996}. The recent star formation in this structure is
likely to be at least partially due to a tidal stream from one of
several recent interactions with other galaxies in the M81 group
\citep{yun1994,yun1999}.  The southeast region of the image is more
crowded as well, where it skims the inner disk.

For measuring the SFH, we divided the field into regions inside the
arm, outside the arm, and the more crowded inner disk, as shown in
Figure~\ref{subdivide}.  Since the crowding and extinction are
different in each of these subregions, we determined the error and
completeness characteristics separately for each subregion.  The final
CMDs for the arm and interarm subregions are shown along with the full
CMD in Figure~\ref{cmds}.

\subsection{Fitting Parameters}\label{binning}

To model a full CMD, several fitting parameters must be chosen.  These
choices include the binary fraction, IMF slope, the area of the CMD to
include in the fit, the approximate distance and mean extinction to
the stars in the field, and the binning of the stellar evolution
models in time and metallicity. Below we discuss how we chose these
parameters and how the choices impact our results.

When populating the model isochrones, we assumed a binary fraction of
0.35 and a \citet{salpeter1955} IMF.  As has been shown by other
studies using this technique \citep[e.g.,][]{williams2007,barker2007},
for photometry that does not include the main sequence of old stars,
the IMF assumed does not affect the relative star formation rates in
the SFH, but does affect the normalization of the SFH.  This
normalization effect occurs because the red clump, RGB, and AGB probe
a narrow range of initial mass.  Therefore changing the slope has
little effect on the relative star formation rates, but a large effect
on the extrapolation of the mass contained in the underlying
unresolved low-mass main-sequence stars.  Since we are not attempting
to determine the precise star formation rate but rather are interested
in the relative SFH within the field, a Salpeter IMF is sufficient for
our purposes.

When fitting the color-magnitude distribution of the stars, we
included all stars brighter than our 60\% completeness limit for the
full field to avoid large completeness corrections in our model
fitting.  This completeness cut corresponds to $m_{F606W} = 29.1$ and
$m_{F814W} = 28.4$, and only stars brighter than these magnitudes were
included in our model fits.  We allowed the distance modulus to range
from 27.75 to 28.2 and the extinction to range from $A_V=0.1$ to
$A_V=0.7$, which allows MATCH to determine the systematic errors that
result from small changes to these parameters and to optimize the
overall CMD fit even in the presence of localized deficiencies in the
model isochrones.  While differential extinction can, in principle, be
a problem for fitting a CMD with a single extinction value, our M81
field lies in a region with very little visible dust structure in {\it
Spitzer} maps \citep{kendall2008}.  Furthermore, modest amounts of
differential extinction have been shown to have little impact on the
results of full CMD fitting \citep{williams2002}.

The best-fitting distances (see Table~\ref{table}) were all consistent
with $(m-M)_0=27.9$ within the errors, which agree with the Cepheid
distance from the {\it HST} key project
\citep[$(m-M)_0=27.8\pm0.2$;][]{freedman1994} and the distance
determined by measuring the tip of the RGB in archival WFPC2
photometry \citep[$(m-M)_0=27.93\pm0.04$;][]{tikhonov2005}.

The best-fitting extinction for the interarm region was consistent
with the value of A$_V$=0.27 obtained from \citet{schlegel1998} for
the Milky Way foreground extinction (see Table~\ref{table}).  We note
that the arm subregion had a measured extinction value that was
significantly higher than the Milky Way foreground, reflecting the
higher dust content expected in the arm region.

\begin{figure}
\centerline{\epsfig{file=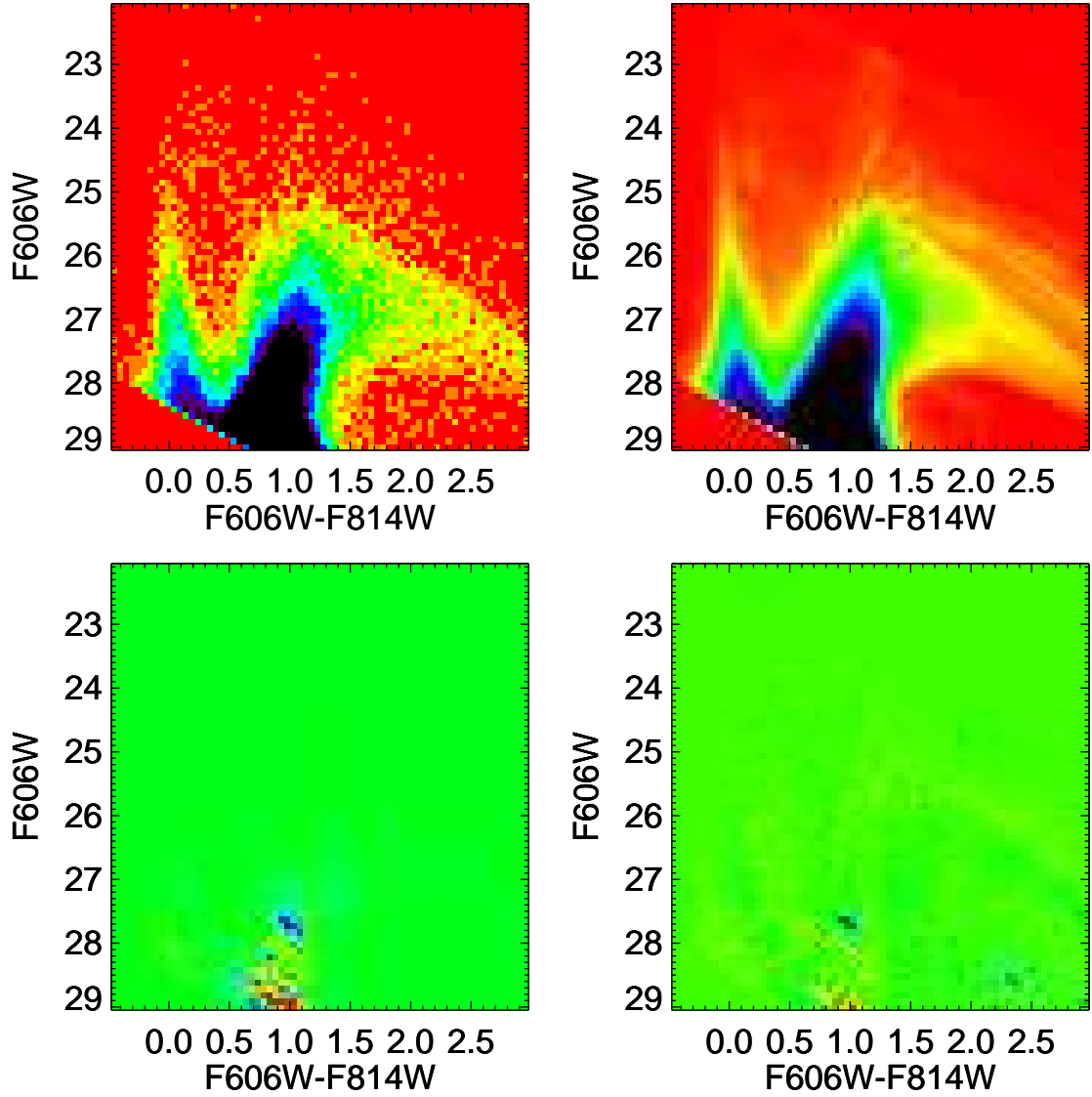,width=6.0in,angle=0}}
\caption{Our best full CMD fit to the data from the entire field.
{\it Upper-left:} The observed CMD. {\it Upper-right:} The
best-fitting model CMD from MATCH. {\it Lower-left:} The residual CMD.
Redder colors denote an overproduction of model stars.  Bluer colors
denote an underproduction of model stars. {\it Lower-right:} The
deviations shown in {\it lower-left} normalized by the Poisson error
in each CMD bin.  This plot shows the significance of the residuals in
{\it lower-left}.  Only the red clump and AGB bump show significant
residuals.}
\label{residuals}
\end{figure}

\begin{figure}
\centerline{\epsfig{file=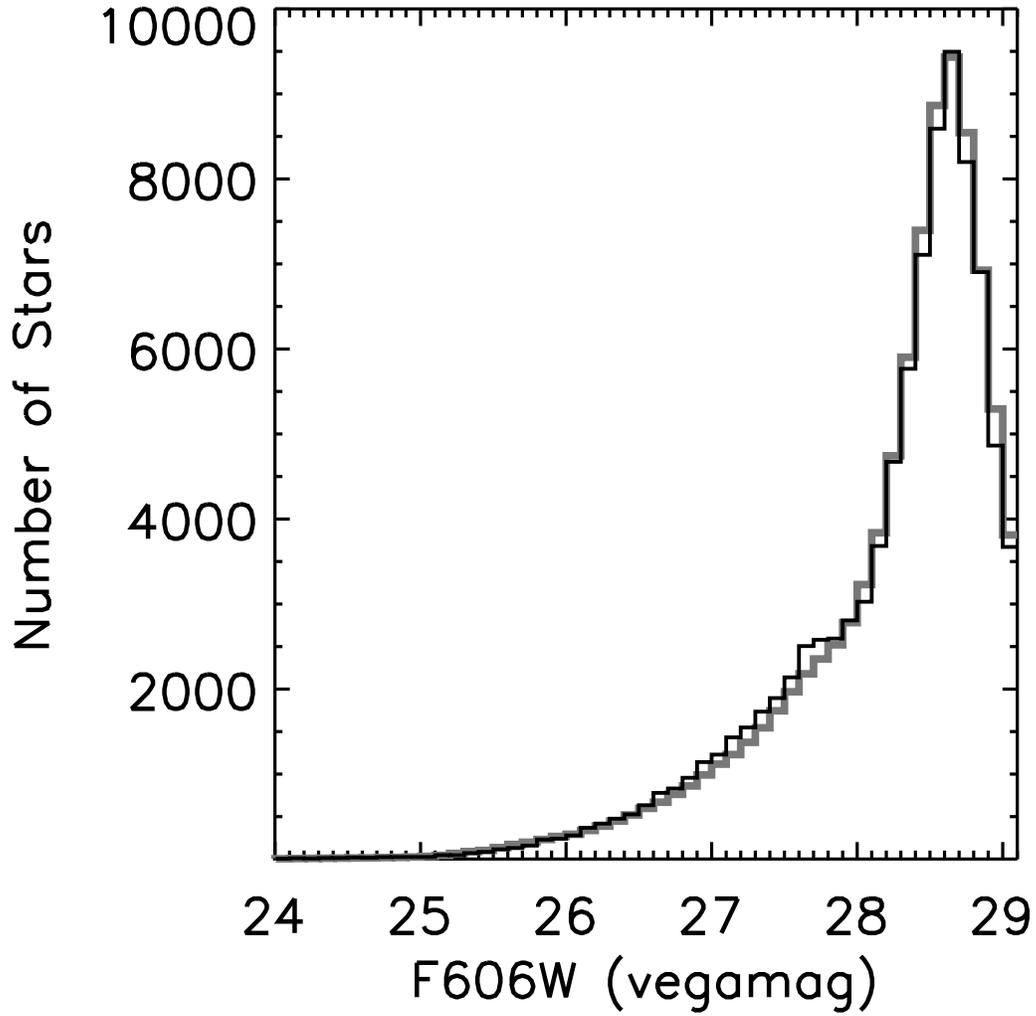,width=6.0in,angle=0}}
\caption{Histograms of the luminosity functions of our best full CMD
fit to the data from the entire field. {\it Black:} The observed
luminosity function of the entire field.  {\it Gray:} The luminosity
function of the best-fit model CMD.}
\label{lf_residuals}
\end{figure}

We used a fine logarithmic time and metallicity resolution (0.1 dex)
when fitting the CMD to allow the best possible fit to the data.  The
fit to the full data set is shown in Figures~\ref{residuals} and
\ref{lf_residuals}.  The full CMD fitting confirms that the models are
not able to perfectly reproduce the red clump and AGB bump, as
expected from our discussion of uncertainties in \S~\ref{rcfit}.
These are the only two features of the CMD that show significant
differences between the data and the best-fitting model (bottom-right
panel of Figure~\ref{residuals}), confirming that the models of these
features still need improvement.  After performing the full CMD fit,
we binned the age and metallicity results to coarser time resolution
to reduce our SFH errors.  Therefore while our fit did not force the
star formation rate or metallicity to be constant within a given
temporal bin, our final SFHs only show the mean rate and metallicity
within each temporal bin to avoid drawing conclusions based on details
of the fit that may not be robust against changes in models and
fitting methods.

\subsection{Error Determination}

In addition to measuring the most likely SFH for our field, we ran
Monte Carlo tests to determine the random uncertainties of the
fits. We generated CMDs by randomly drawing stars from our observed
CMD, allowing each star to be drawn any number of times.  We then
measured SFHs for the resulting CMDs and calculated differences from
the best-fit to the actual data.  We generated 100 samples with the
same number of stars as our observed CMD, adopting each of the SFHs
measured from our data to determine the random errors for each
subfield.  These errors were then added in quadrature to the
systematic errors, determined by fitting the CMDs with a range of
possible distance and reddening values, to provide the our final
errors on the rate, metallicity, and cumulative fraction of stars
formed as a function of time.

We note that our choice of time bin widths was very coarse for ages
$>$2~Gyr to reflect our sensitivity to age from the RGB and AGB.  From
$\sim$2--8~Gyr (9.3$<$log(age)$<$9.9), the age distribution comes
mostly from the relatively small number of stars on the AGB, which is
a very short-lived and difficult to model stage of evolution.  From
8--14~Gyr (9.9$<$log(age)$<$10.15) age has very little effect on the
morphology of the AGB and RGB features.  In addition, our metallicity
measurements at young ages ($<$100 Myr) have large errors.  Our only
metallicity information for these ages comes from the relatively small
number of stars on the short-lived He-burning sequences.  Despite
these unavoidable sources of uncertainty in age and metallicity,
overall we are able to obtain very reliable estimates of the relative
contributions of stars of old ($>$8~Gyr), intermediate (2--8~Gyr), and
young ($<$2~Gyr) ages. Furthermore, we obtain reliable metallicities
covering all but the youngest ages.  Finally, the cumulative age
distribution is stable against uncertainties at intermediate ages
because systematic errors in the star formation rates in adjacent time
bins are typically anti-correlated.

\subsection{The SFH of the Full Field} 

\begin{figure}
\centerline{\epsfig{file=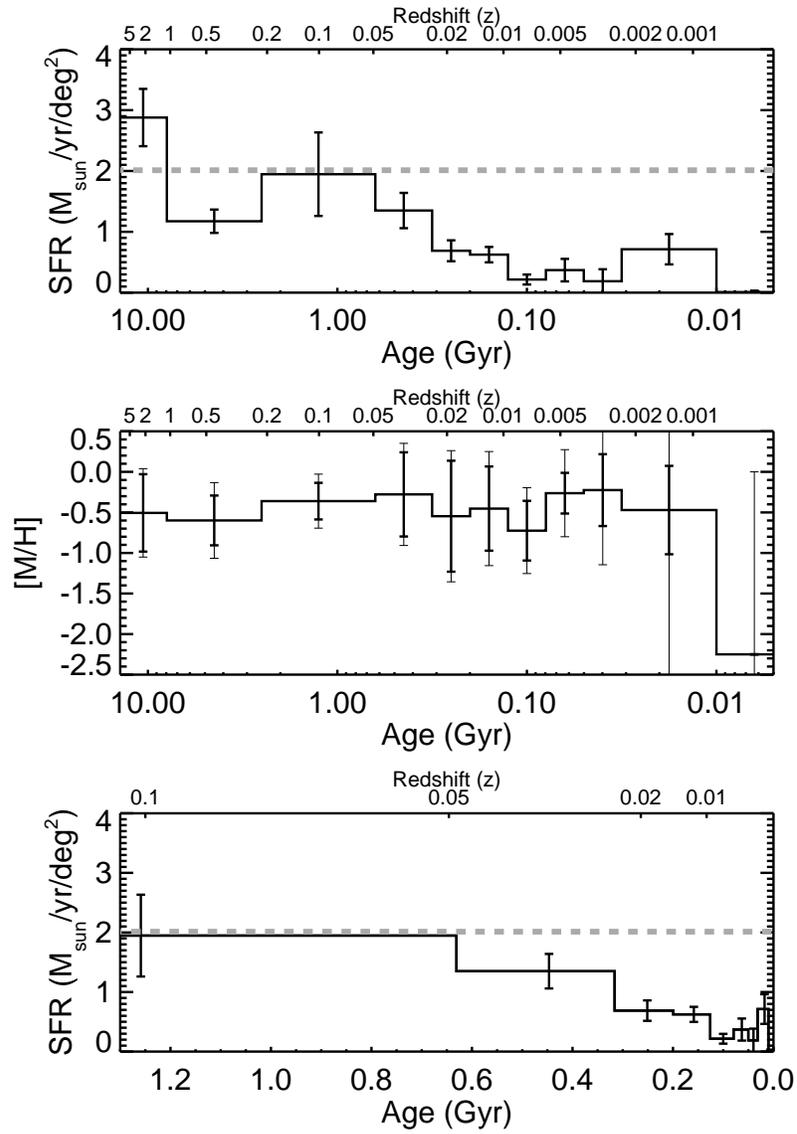,height=6.0in,angle=0}}
\caption{The SFH of the entire ACS field as determined by the MATCH
package. {\it Top:} The solid histogram marks the star formation rate
(normalized by sky area) as a function of time for the past 14
Gyr. The dashed line marks the best-fitting constant star formation
rate model. {\it Middle:} The mean metallicity and metallicity range
of the population as a function of time.  Heavy error bars mark the
measured metallicity range, and lighter error bars mark how that range
can slide because of errors in the mean metallicity. {\it Bottom:}
Same as {\it top}, but showing only the results for the past 1.3 Gyr.}
\label{all_sfh}
\end{figure}

\begin{figure}
\centerline{\epsfig{file=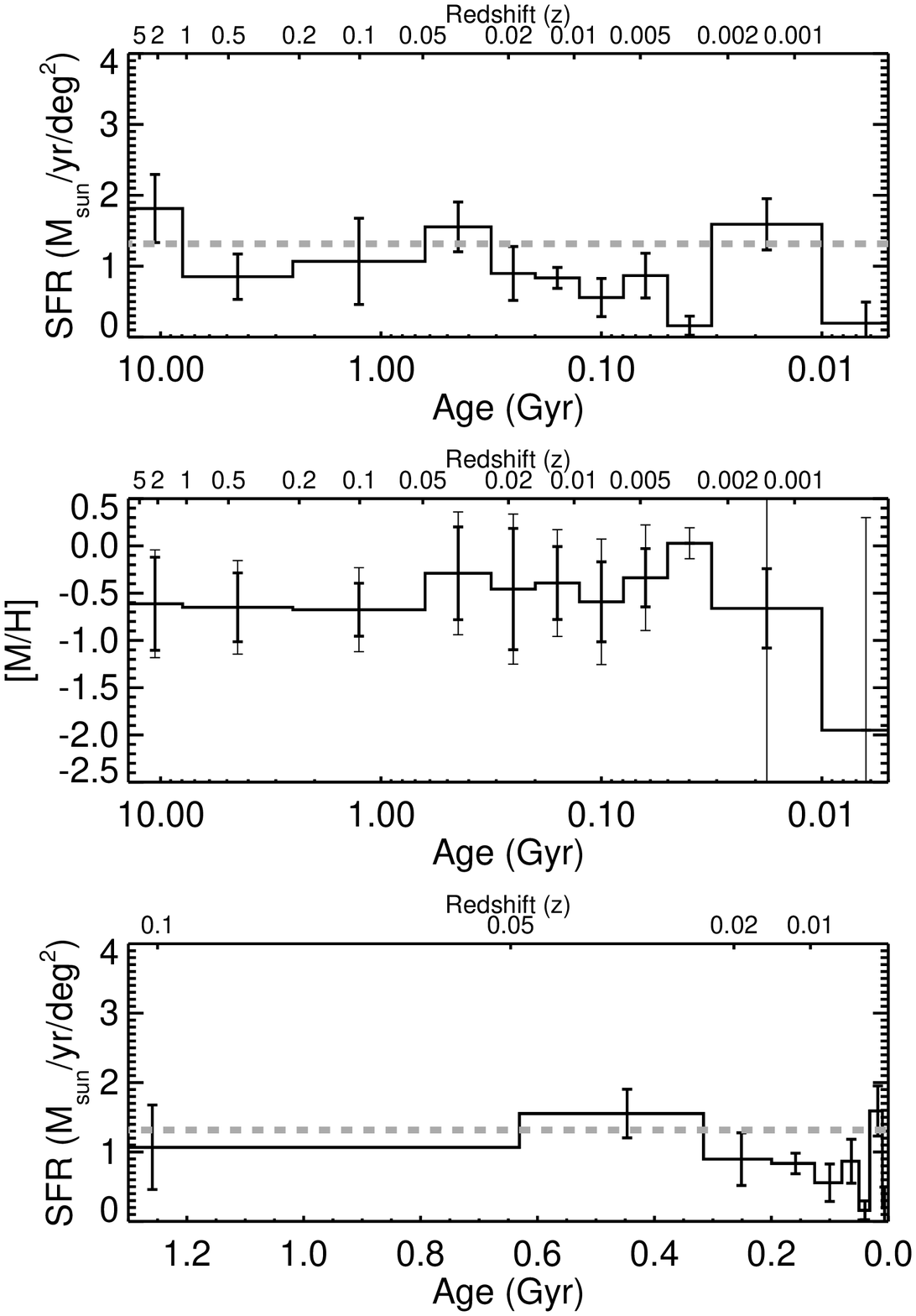,height=6.0in,angle=0}}
\caption{The SFH of the arm subregion as determined by the MATCH
package. Lines, error bars, and panels are the same as in
Figure~\ref{all_sfh}.}
\label{arm_sfh}
\end{figure}

\begin{figure}
\centerline{\epsfig{file=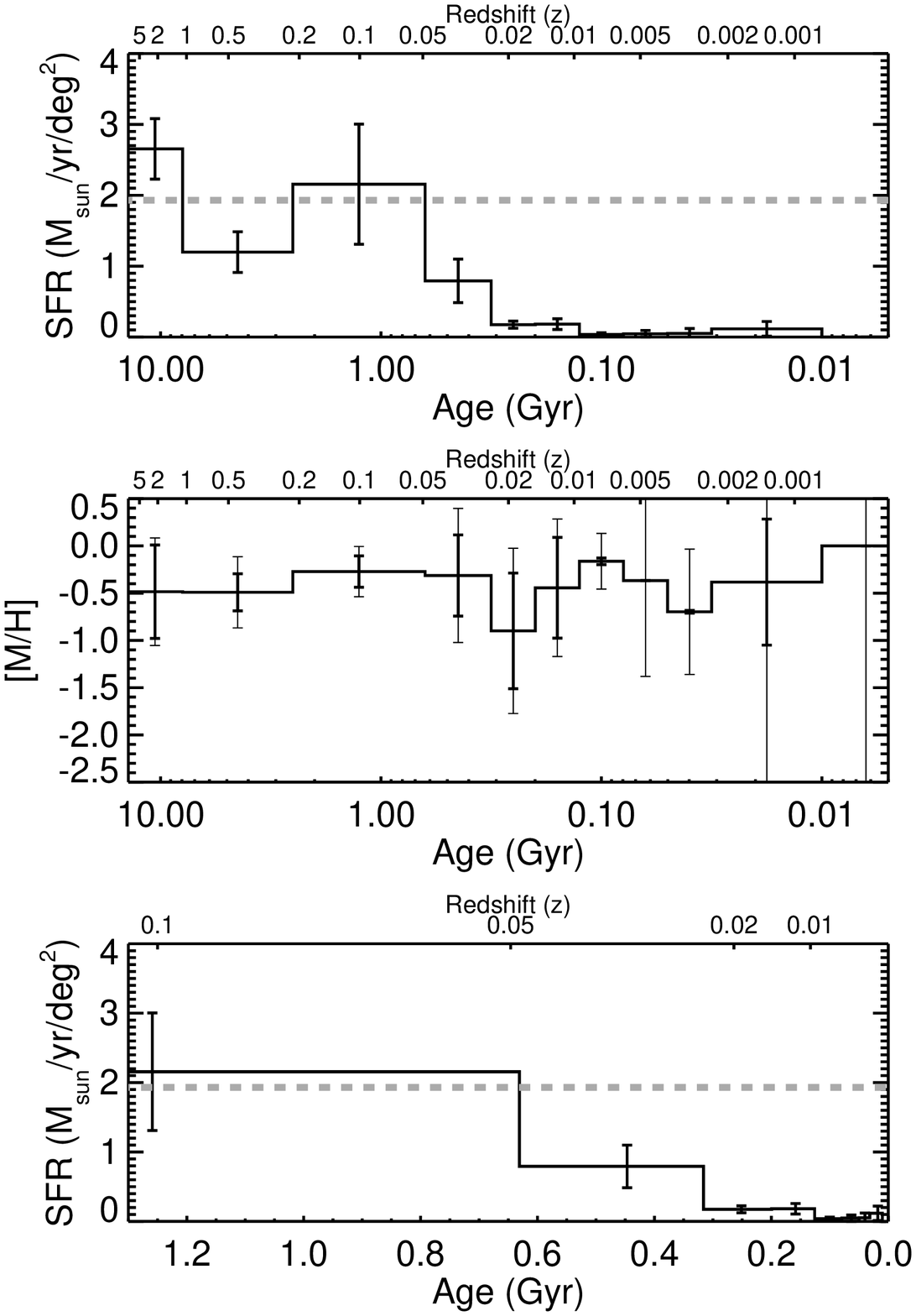,height=6.0in,angle=0}}
\caption{The SFH of the interarm subregion as determined by the MATCH
package. Lines, error bars, and panels are the same as in
Figure~\ref{all_sfh}.}
\label{noarm_sfh}
\end{figure}

\begin{figure}
\centerline{\epsfig{file=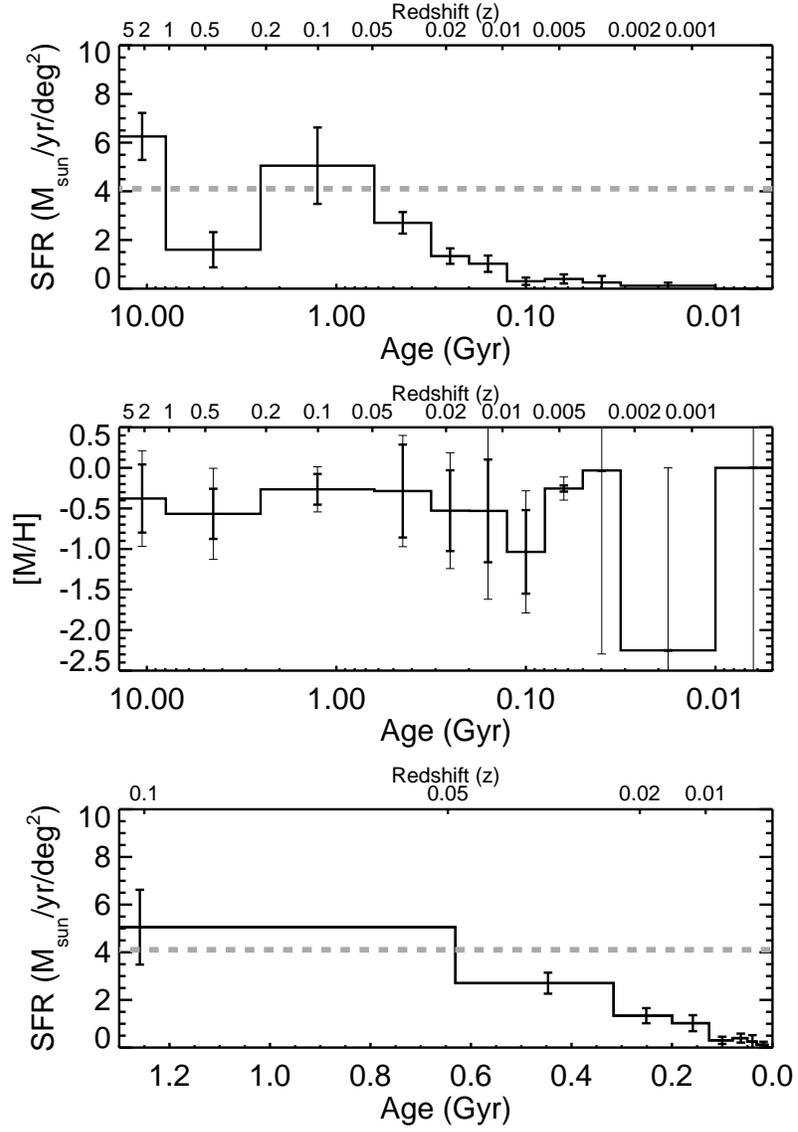,height=6.0in,angle=0}}
\caption{The SFH of the crowded southern subregion as determined by
the MATCH package. Lines, error bars, and panels are the same as in
Figure~\ref{all_sfh}.}
\label{crowd_sfh}
\end{figure}

\begin{figure}
\centerline{\epsfig{file=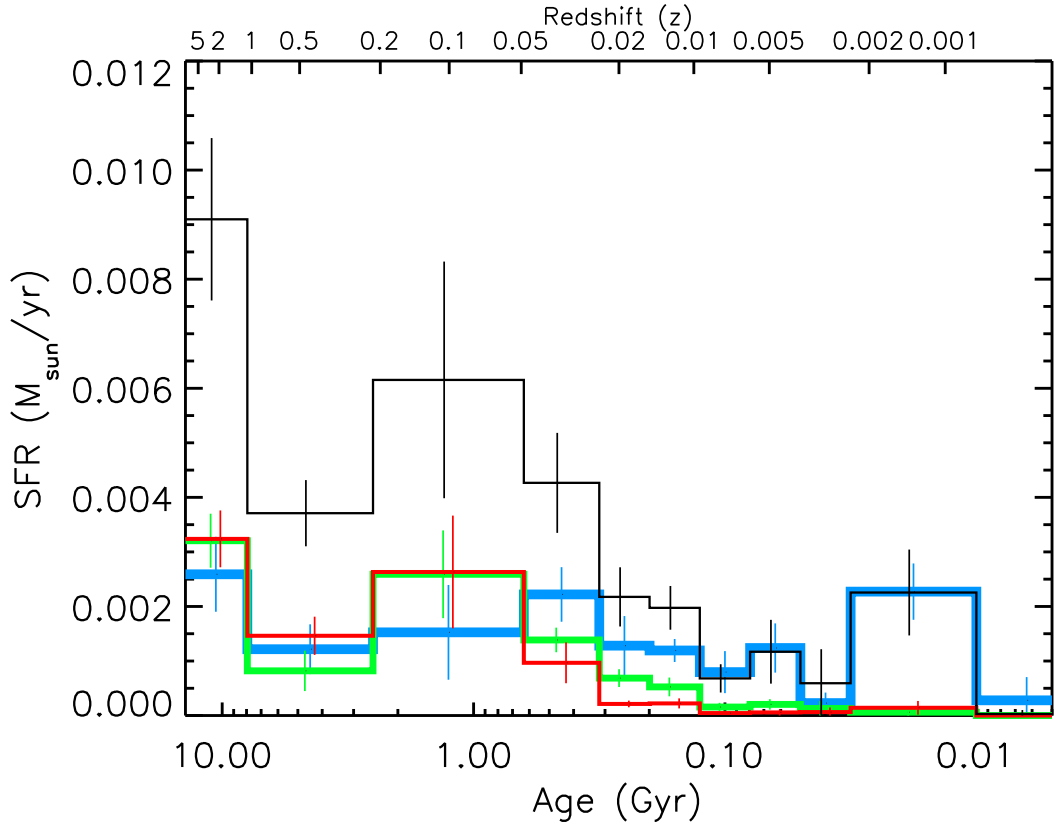,width=6.0in,angle=0}}
\caption{The SFH of all subregions are overplotted.  Rates are not
normalized by area in order to show the contribution of each subregion
to the star formation of the total field.  {\it Black:} The SFH of the
full field as determined by the MATCH package. {\it Blue:} The SFH of
the arm subregion as determined by the MATCH package.  {\it Green:}
The SFH of the crowded, southern subregion as determined by the MATCH
package.  {\it Red:} The SFH of the interarm subregion as determined
by the MATCH package.}
\label{comparesfh}
\end{figure}

\begin{figure}
\centerline{\epsfig{file=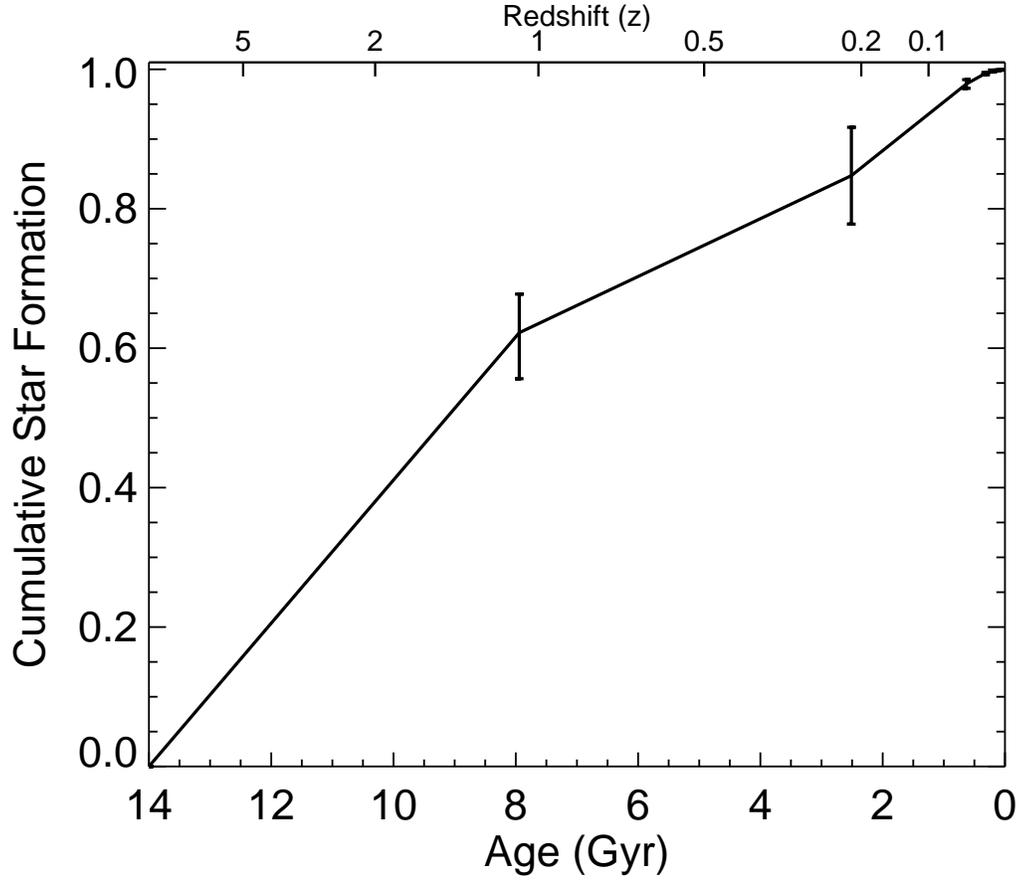,width=6.0in,angle=0}}
\caption{{\it Black:} The normalized cumulative star formation of the
full field as determined by the MATCH package.}
\label{cum}
\end{figure}

The SFH of the entire field and the three independent subfields are
shown in Figures~\ref{all_sfh}--\ref{cum}.  For the history of star
formation in the full field (Figure~\ref{all_sfh} and
Table~\ref{table2}), we find that more than 50\% of the stars
currently at 5 scale lengths from the galactic center formed by $z\sim
1$, and $\sim$70\% formed by $z\sim 0.5$. The bulk of the stars have
$-1\lap {\rm [M/H]} \lap 0$, with no significant metallicity
difference between the arm and interarm regions and with no evidence
for a significant metal-poor component.  Such a metallicity is
consistent with the mean metallicities found by \citet[][$<${{\rm
[M/H]}}$> \sim -0.65$]{tikhonov2005} from stellar photometry in 5
WFPC2 fields spanning a range of galactocentric distances out to
$\sim$5 scale lengths.  It therefore appears that M81 was chemically
enriched very early in its history.  Moreover, the metallicity has not
changed by more than $\sim$1.0 dex (and possibly as little as 0.5 dex)
since its very early history.

Although our metallicity results for recent times have large errors,
their values (-0.5$<{\rm [M/H]}<$0.0 at ages from 10 Myr to 100~Myr;
Figure~\ref{all_sfh}) are consistent with the gas-phase metallicity at
this radius (${\rm [O/H]}\sim-0.3$) inferred from the abundance
gradients of \citet{zaritsky1994}, suggesting that the gas responsible
for this structure is from M81 and not recently accreted gas from a
more metal-poor interacting satellite.  However, this metallicity does
not rule out the possibility that the gas came from a metal-rich
satellite, such as NGC 3077 \citep[{{\rm
[M/H]}}$\sim$0,][]{martin1997}.

\subsection{The SFHs of the Subregions} 

Figure~\ref{comparesfh} shows the SFHs of the three subregions
overplotted with the SFH of the total field.  It is clear that the sum
of the results of the three subregions measured independently is
consistent with the result of the entire field measured at once.  The
consistency between the 3 independently measured SFHs and the total
SFH for the field confirms that our measurement technique provides
reliable and consistent results in a field containing regions with
moderately different crowding and extinction properties.  Furthermore,
the more crowded portion of the field, closest to the galaxy center,
has a SFH more similar to that of the interarm region than to that of
the arm region, showing that this area is not part of an arm and is of
higher density only because it is closer to the galaxy center.

While these three regions should have obvious differences in their
recent SFHs (based on the distribution of main sequence stars;
cf. Figure~\ref{subdivide}), the older stellar populations should be
well mixed on timescales longer than the dynamical time ($\gap$0.5
Gyr).  Indeed, we measure similar SFHs at times $\gap$0.5 Gyr.  At
younger ages, the interarm region shows a clear deficit of stars
younger than a few hundred Myr.  Fractionally, this difference
corresponds to $\sim$2\% of the total star formation (see
Figure~\ref{cum} and Table~\ref{table2}).  The arm region of our field
contains essentially all of the stars with ages $\lap$100 Myr.  This
shows that the stars have not scattered out of the arm structure on
this timescale.  In addition, there is a significantly higher fraction
of stars with ages $\lap$0.5~Gyr in the arm region than in the other
regions.  It therefore appears that the arm structure in this region
has a lifetime of $\gap$100 Myr and that full dispersion occurs on a
timescale of $\sim$0.5 Gyr.  The characteristic width of the arm
region, divided by the typical lifetime gives an approximate speed for
the diffusion of stars. Assuming an arm width of $\sim$1 kpc
\citep{westpfahl1998} divided by a lifetime of $\sim$100 Myr yields a
diffusion speed of $\sim$10 km s$^{-1}$, which is similar to the
velocity dispersion of B stars in the Milky Way disk
\citep{dehnen1998}.  Therefore the timescales we are measuring for
this arm feature are consistent with the stellar kinematics of our own
Galactic disk. Since our data contain main-sequence and He-burning
sequence information for the populations of these ages, spatially
resolved SFHs, with a time resolution of $\sim$ 25 Myr over the last
$\sim$ 300 Myr are currently in progress and will be presented in
future papers (\citealp{gogarten2008}; B. Williams et al.,
in preparation).

\section{Conclusions}

We have performed deep resolved stellar photometry of an ACS field in
the outer disk of M81. The metallicities of the stars in the field
appear to have increased by at most $\sim$0.5 dex over the past 10 Gyr
from -1.0$\lap$[M/H]$\lap$-0.5 to -0.5$\lap$[M/H]$\lap$0, suggesting
early enrichment and a mechanism at work diluting the enrichment
products of roughly continuous star formation.

\begin{figure}
\centerline{\epsfig{file=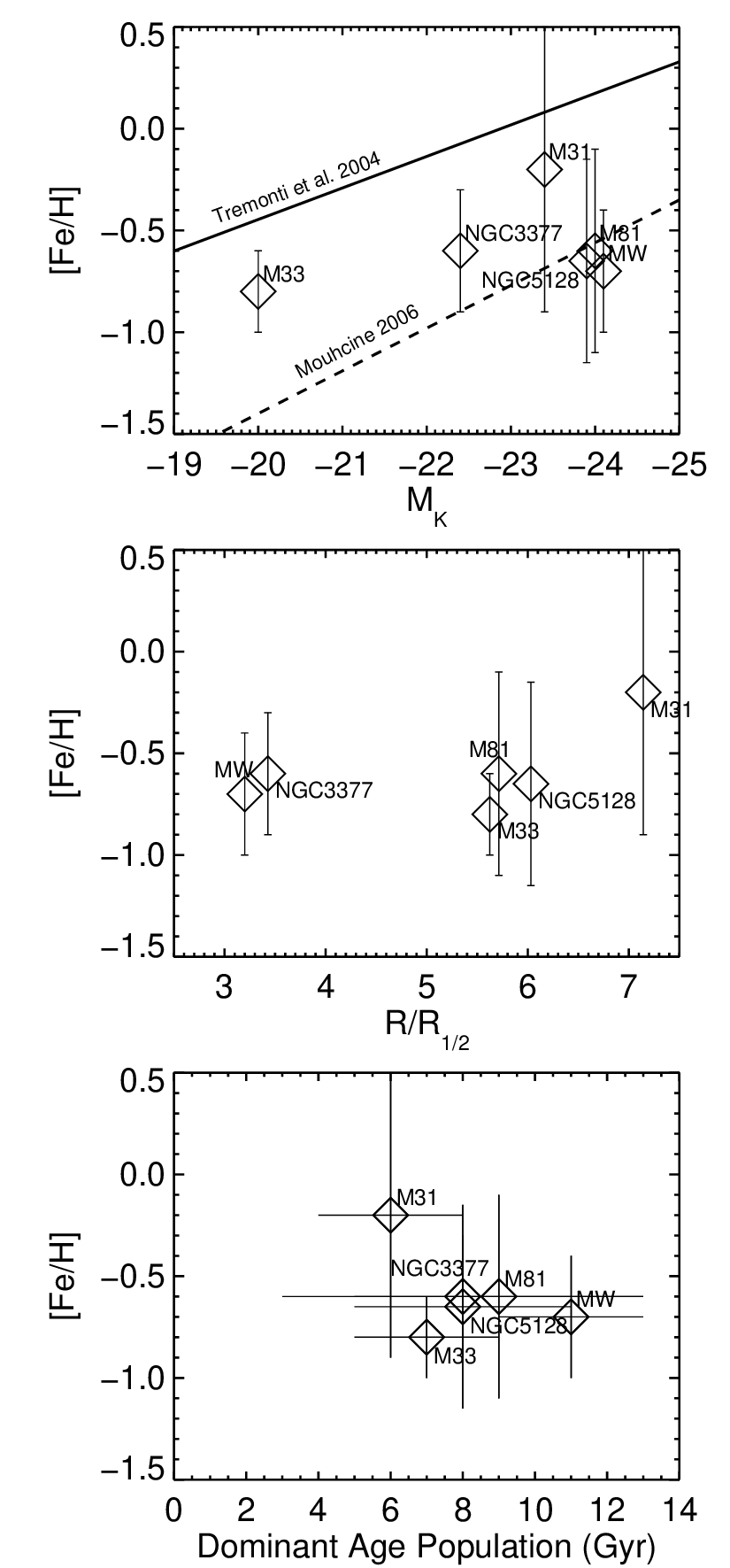,height=6.5in,angle=0}}
\caption{\footnotesize Metallicity range of the dominant stellar
populations in deep resolved photometry for M31 \citep{brown2006}, M33
\citep{barker2007}, M81 (this work), NGC 5128 \citep{rejkuba2005}, and
NGC 3377 \citep{harris2007} are plotted along with that of the Milky
Way thick disk \citep{allendeprieto2006} against several other
properties. {\it Top:} The metallicities as a function of the absolute
K-band magnitude of the galaxy \citep{skrutskie2006}.  The solid and
dashed lines show the luminosity-metallicity relations determined by
\citet[][gas phase metallicity of galaxy central
regions]{tremonti2004} and \citet[][stellar red peak metallicities of
galaxy ``halos'']{mouhcine2006}, respectively.  These relations were
converted from B-band and V-band to K-band using the Tully-Fisher
calibrations of \citet{verheijen2001} and \citet{sakai2000}.  The
Milky Way luminosity was calculated by applying $V_{rot}$= 220 km
s$^{-1}$ to the Tully-Fisher calibration of
\citet{verheijen2001}. {\it Middle:} The populations' metallicities as
a function of the radii at which they were sampled (normalized to the
half-light radius of the galaxy).  {\it Bottom:} The populations'
metallicities as a function of their ages.}
\label{galaxies}
\end{figure}

Similar behavior has been seen in the thick disk of the Milky Way and
several other nearby large galaxies.  Figure~\ref{galaxies} shows the
metallicities of the dominant stellar population for several nearby
galaxies and the Milky Way thick disk as a function of galaxy
luminosity, field location, and dominant stellar age.  The Milky Way
thick disk has characteristic ages of 9--12 Gyr and metallicities of
[M/H]$\gap -0.7$ \citep{gilmore1995,prochaska2000,allendeprieto2006}.
Thick disks in somewhat smaller galaxies also appear to have
metallicities $>$-1.0 \citep{seth2005}.  In addition, deep {\it HST}
photometry of the M31 disk has revealed populations dominated by stars
in this age and metallicity range
\citep{williams2002,olsen2006,brown2006}, as has deep photometry of
the outer disk of M33 \citep{barker2007}.  All of these results
suggest that the histories of these large disk galaxies may have been
similar.

Furthermore, even in the outskirts of large ellipticals, such as
NGC~5128 \citep[${\rm [M/H]}\sim$-0.6 and age$\sim$8
Gyr,][]{rejkuba2005} and NGC~3377 \citep[${\rm [M/H]}\sim$-0.6 and age
$>$3 Gyr,]{harris2007}, the age and metallicity of the dominating
population is similar to that seen in the outer regions of M81 and
other large disks. Taken together, these data point to a large galaxy
formation scenario with rapid early enrichment to [M/H]$\gap$-1.0,
before lookback times of $\sim$7 Gyr ($z>0.9$).

There is also significant evidence that most of the stars in this
field were formed by $z\sim 1$.  Such a result is in general agreement
with the findings of several recent galaxy surveys, which find that
the disk galaxy population appears to have undergone little growth
since $z\sim 1$ \citep[e.g.,
][]{melbourne2007,sargent2007,papovich2005,barden2005,ravindranath2004,lilly1998}.
While M81 is only one large disk galaxy and our field is only a small
portion of it, the similarity to Milky Way, M31, and even lower mass
M33's stellar populations supports the same scenario.  Measurements of
the stellar populations of local disks therefore strengthen the
results of some galaxy surveys by attacking the problem with a
completely independent technique and finding a similar result.

Finally, the spatial distribution of main-sequence stars in the field
show that the field partially covers the outskirts of a spiral arm as
well as an interarm region. Detailed analysis of the resulting CMDs
for the contrasting regions shows that the difference between the
populations is due to the fraction of stars with ages younger than
$\sim$0.5~Gyr ago and results from only a small percentage ($\sim$2\%)
of the stars, similar to the arm and interarm populations seen in M31
disk populations \citep{williams2002}.  In addition, stars younger
than $\sim$100 Myr appear to be confined to the arm region, suggesting
the structure survives for at least 100 Myr.

Support for this work was provided by NASA through grant GO-10915 from
the Space Telescope Science Institute, which is operated by the
Association of Universities for Research in Astronomy, Incorporated,
under NASA contract NAS5-26555.

\clearpage

\begin{deluxetable}{lcc}\tablewidth{7cm}
\tablecaption{Distances and Extinction Values from MATCH}
\tablehead{
\colhead{Subfield} &
\colhead{$(m-M)_0$} &
\colhead{$A_V$} 
}

\startdata
Full & 27.89$\pm$0.05  & 0.38$\pm$0.06\\
Arm & 27.88$\pm$0.08  & 0.45$\pm$0.06\\
Interarm & 27.92$\pm$0.06 & 0.33$\pm$0.06\\
Crowd & 27.88$\pm$0.06 & 0.32$\pm$0.06\\
\enddata
\label{table}
\end{deluxetable}

\begin{deluxetable}{lcc}\tablewidth{7cm}
\tablecaption{Cumulative Fraction of Stars Formed at Each Age}
\tablehead{
\colhead{Redshift} &
\colhead{Age (Gyr)} &
\colhead{Fraction of Stars Formed} 
}

\startdata
0.001 & 0.01&1.000$^{+0.000}_{-0.000}$\\
0.002 & 0.03&0.999$^{+0.000}_{-0.000}$\\
0.004 & 0.05&0.999$^{+0.000}_{-0.000}$\\
0.006 & 0.08&0.999$^{+0.000}_{-0.000}$\\
0.010 & 0.13&0.999$^{+0.000}_{-0.001}$\\
0.015 & 0.20&0.997$^{+0.001}_{-0.001}$\\
0.024 & 0.32&0.994$^{+0.001}_{-0.002}$\\
0.049 & 0.63&0.979$^{+0.006}_{-0.006}$\\
0.216 & 2.51&0.848$^{+0.069}_{-0.070}$\\
1.080 & 7.94&0.622$^{+0.055}_{-0.066}$\\
\nodata & 14.1&0.000$^{+0.000}_{-0.000}$\\
\enddata
\label{table2}
\end{deluxetable}

\clearpage

\clearpage

%\begin{landscape}

%\end{landscape}

\clearpage

\end{document}